%% file: main.tex
\documentclass[fleqn,10pt]{wlscirep}
\usepackage[utf8]{inputenc}
\usepackage[T1]{fontenc}
\usepackage{subcaption}

\title{A Global Commuting Origin-Destination Flow Dataset for Urban Sustainable Development}

\author[1,2]{Can Rong}
\author[1,2]{Jingtao Ding}
\author[3]{Meng Li}
\author[1,2,*]{Yong Li}
\affil[1]{Department of Electronic Engineering, Tsinghua University, Beijing, P.R. China}
\affil[2]{Beijing National Research Center for Information Science and Technology (BNRist), Beijing, P.R. China}
\affil[3]{Department of Civil Engineering, Tsinghua University, Beijing, P.R. China}
\affil[*]{corresponding author(s): Yong Li (liyong07@tsinghua.edu.cn)}

\begin{abstract} 
Commuting Origin-Destination~(OD) flows capture movements of people from residences to workplaces, representing the predominant form of intra-city mobility and serving as a critical reference for understanding urban dynamics and supporting sustainable policies. However, acquiring such data requires costly, time-consuming censuses. In this study, we introduce a commuting OD flow dataset for cities around the world, spanning 6 continents, 179 countries, and 1,625 cities, providing unprecedented coverage of dynamics under diverse urban environments. Specifically, we collected fine-grained demographic data, satellite imagery, and points of interest~(POIs) for each city as foundational inputs to characterize the functional roles of urban regions. Leveraging these, a deep generative model is employed to capture the complex relationships between urban geospatial features and human mobility, enabling the generation of commuting OD flows between urban regions. Comprehensively, validation shows that the spatial distributions of the generated flows closely align with real-world observations. We believe this dataset offers a valuable resource for advancing sustainable urban development research in urban science, data science, transportation engineering, and related fields.
\end{abstract}
\begin{document}

\flushbottom
\maketitle

\thispagestyle{empty}


\input{Sec_1intro}
\input{Sec_2method}
\input{Sec_3data}
\input{Sec_4eval}

\section*{Usage Notes}
\begin{figure}[ht]
    \centering
    \includegraphics[width=0.6\linewidth]{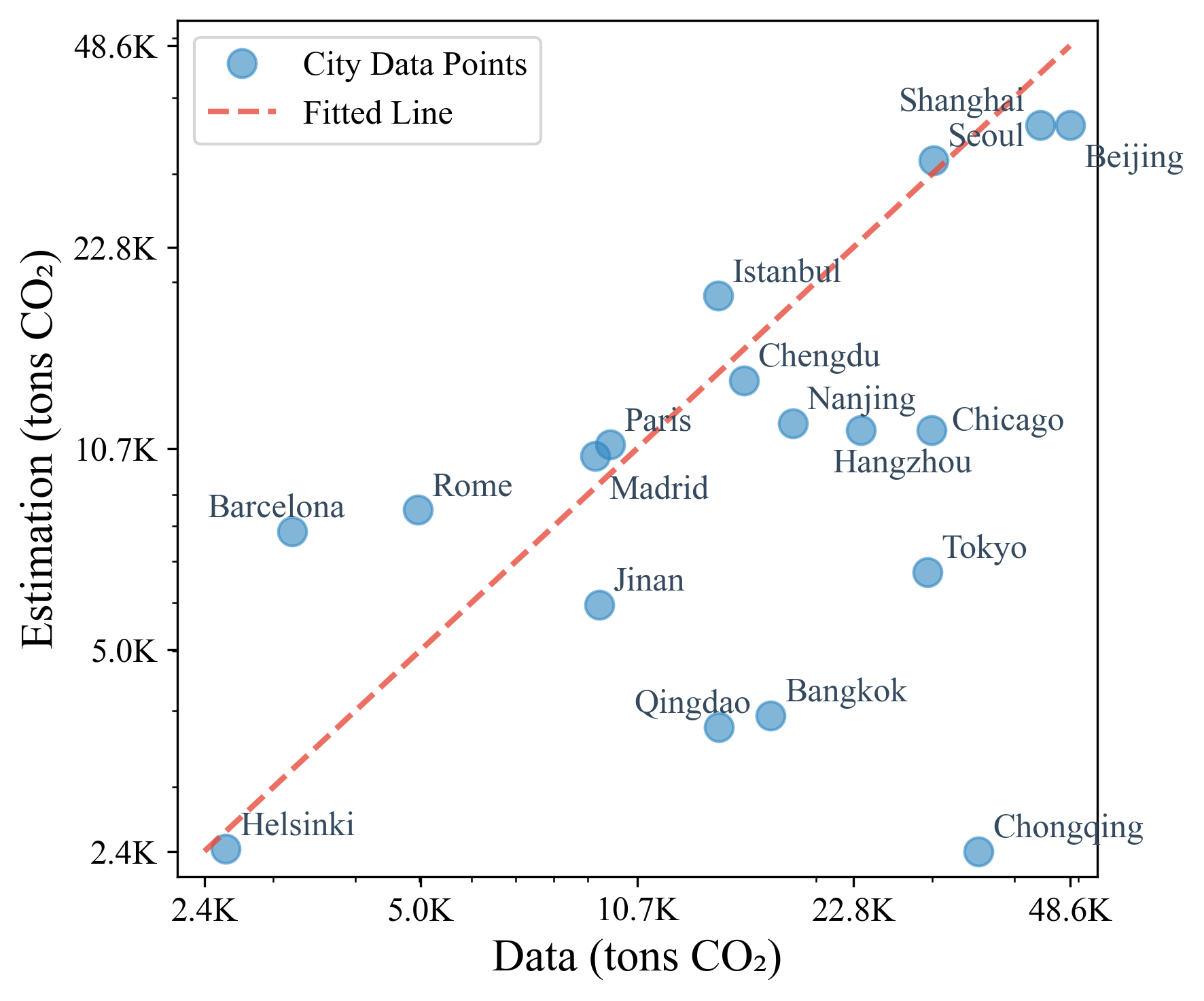}
    \caption{Comparison between estimated and measured urban transportation carbon emissions.~(Downstream task to achieve \textbf{SDGs-9,13})}
    \label{fig:carbon_estimation}
\end{figure}

\begin{figure}[ht]
    \centering
    \includegraphics[width=\linewidth]{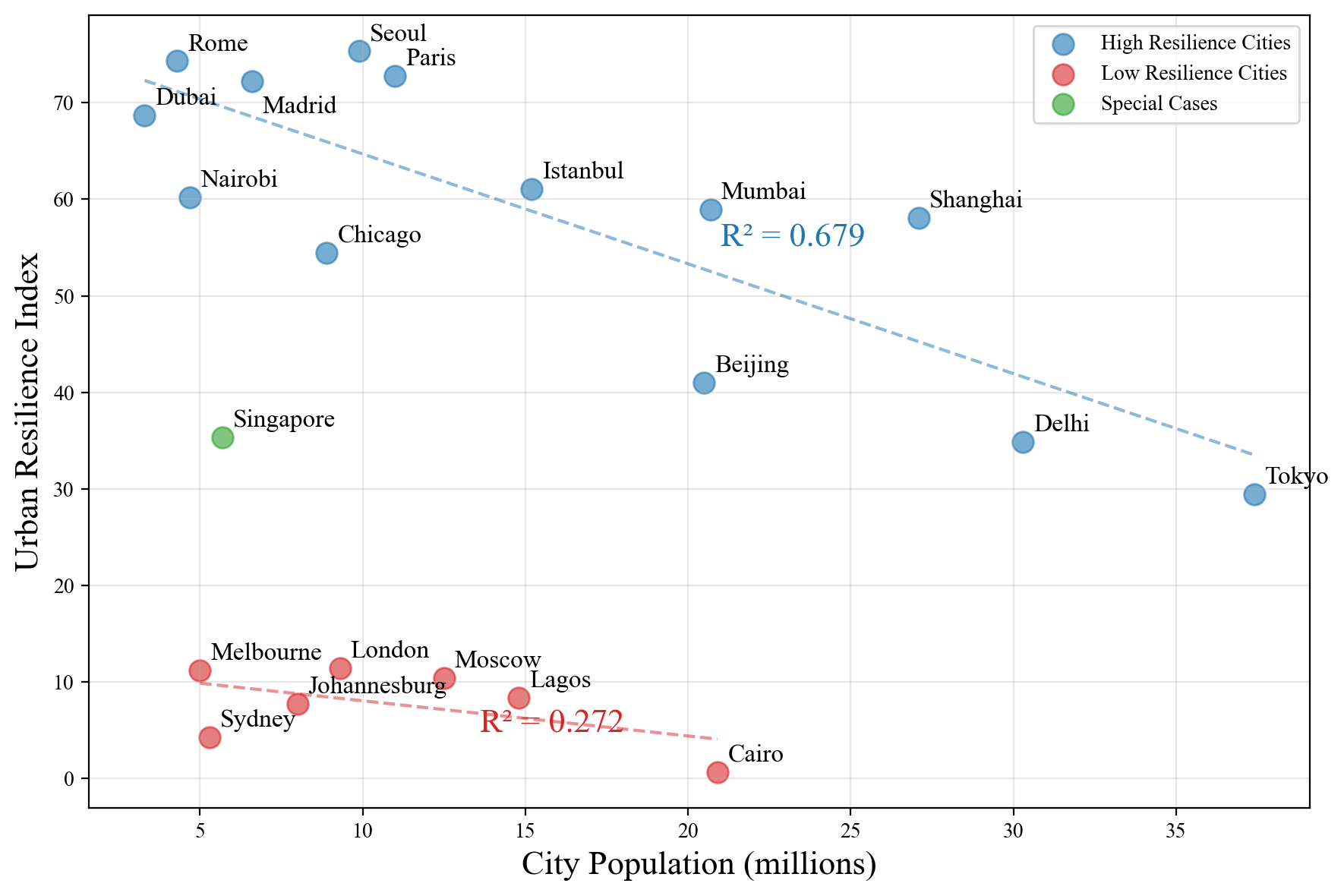}
    \caption{Resilience analysis of commuting OD networks.~(Downstream task to achieve \textbf{SDGs-9,11})}
    \label{fig:resilience_groups_analysis}
\end{figure}

This dataset provides intra-city commuting OD flows for 1,625 cities worldwide. It spans a wide range of urban environments, cultural contexts, development stages, and infrastructural characteristics, making it an invaluable resource for advancing research on urban sustainable development in the following aspects:
\begin{itemize}[leftmargin=*]
    \item \textbf{Urban Planning:} The spatial arrangement of functional zones, such as residential, commercial, and industrial areas, plays a pivotal role in shaping commuting patterns, with significant implications for urban sustainability. Inefficient urban planning often leads to excessive long-distance commutes, undermining operational efficiency, increasing travel costs, and intensifying environmental burdens such as greenhouse gas emissions and air pollution~\cite{lin2024review,raudsepp2021long,hamilton1982wasteful}. Commuting OD flows provide a valuable analytical tool for evaluating commuting efficiency and diagnosing spatial mismatches in urban planning~\cite{hamilton1982wasteful,sandow2010persevering,mitra2019they}. By quantifying and visualizing these flows, policymakers and urban planners can identify inefficiencies, optimize zoning strategies, and promote sustainable mobility solutions, such as public transit-oriented development or mixed-use neighborhoods. As shown in Figure~\ref{fig:resilience_groups_analysis}, different urban planning approaches lead to varying levels of urban resilience. Although cities tend to become more vulnerable as they grow larger, reliable planning can enhance their resilience through optimized spatial organization and efficient transportation networks.
    \item \textbf{Transportation Engineering:} OD flows, representing the travel demand of citizens, serve as a foundational input for transportation simulation and analysis~\cite{sakai2020simmobility,adnan2016simmobility,de2024modelling}. This facilitates evidence-based decision-making to optimize transportation infrastructure, reduce congestion, and support sustainable urban mobility. Furthermore, commuting OD flows can guide the integration of green transportation solutions~\cite{cantwell2009examining,chng2016commuting}, such as public transit networks and active mobility options, thereby contributing to the reduction of energy consumption, emissions, and environmental impacts in urban areas.
    \item \textbf{Public Health:} Population movement and aggregation in cities are key drivers of infectious disease spread~\cite{balcan2009multiscale}. Understanding urban mobility, particularly commuting OD flows, offers critical insights for outbreak tracing and control~\cite{balcan2009multiscale,mistry2021inferring}. As a common form of daily mobility, commuting flows enable network analysis to identify high-risk hubs or neighboring nodes~\cite{hao2020understanding} for targeted interventions, such as surveillance, resource allocation~\cite{chen2022strategic}, or mobility restrictions~\cite{kraemer2020effect}. Utilizing OD flow data aligns public health strategies with sustainable urban development, fostering resilience and disease preparedness.
    \item \textbf{Energy Use and Environmental Protection:} Daily commuting leads to substantial travel costs, carbon emissions, and pollutants. OD flows can identify high-density commuting corridors, enabling the strategic deployment of renewable energy-powered transit systems, such as electric buses or metro networks, to reduce fossil fuel reliance~\cite{modarres2017commuting}. Additionally, OD insights support urban energy planning by optimizing the placement of electric vehicle charging stations along key routes and prioritizing energy-efficient building retrofits in high-footfall zones~\cite{zhang2022cross,xu2018planning,morfeldt2022impacts}. These targeted strategies help balance energy demand, reduce emissions, and promote environmental sustainability. As shown in Figure~\ref{fig:carbon_estimation}, commuting OD flows enable accurate estimation of urban transportation carbon emissions, providing valuable insights for environmental protection and sustainable development policy-making.
\end{itemize}

It is important to acknowledge certain limitations of this dataset. First, the dataset exclusively focuses on intra-city commuting OD flows, meaning that inter-city commuting flows are not included. Second, the dataset captures only commuting OD flows during work hours, excluding other types of trips such as shopping, leisure, and other activities. Third, the dataset represents static commuting OD flows and does not account for temporal variations. While differences in development stages across cities may serve as a limited proxy for temporal variations, this approach lacks the precision and granularity offered by city-specific historical development data, which remains a significant limitation.

For datasets containing multiple file types, we recommend using Python for loading and processing the data. Specifically, GeoPandas is ideal for loading city boundaries and region division data provided in Shapefile format, while NumPy is well-suited for loading the commuting OD flow data. Additionally, the Shapefile data can be loaded using GIS software such as ArcGIS. Python, integrated within ArcGIS, can also be used to process the commuting OD flow data, enabling seamless handling of both data types in a geospatial context.

To facilitate easy access for researchers across different domains, we also provide a web-based graphical interface at \url{https://fi.ee.tsinghua.edu.cn/worldod/} that allows users to generate commuting OD flows for any region of interest without writing code.

\section*{Code availability}
The Python codes that give the examples for loading and processing the data are available at the Github repository \url{https://github.com/tsinghua-fib-lab/WorldCommuting-OD}. The environment requirements are provided in the repository.

\bibliography{reference}

\section*{Acknowledgements}
In this work, we collect the data from multiple sources, including Who's On First, WorldPop, Esri World Imagery, and OpenStreetMap. We would like to express our gratitude to the contributors of these datasets. We also thank the advancing large multi-modal model, such as RemoteCLIP, for providing the convenient usage of its pre-trained models as a key component in our dataset construction pipeline. 

\section*{Author contributions statement}

Yong Li and Jingtao Ding conceived the idea of the dataset and designed the dataset construction pipeline. Can Rong wrote the code for the generation including: downloading and processing the data from WorldPop, Esri World Imagery, and OpenStreetMap, training the graph diffusion model, and generating the commuting OD flows. Can Rong, Jingtao Ding, and Yong Li wrote the manuscript. All authors reviewed the manuscript.

\section*{Competing interests} (mandatory statement) 

The corresponding author is responsible for providing a \href{https://www.nature.com/sdata/policies/editorial-and-publishing-policies#competing}{competing interests statement} on behalf of all authors of the paper. This statement must be included in the submitted article file.

\section*{Figures \& Tables}

\begin{figure}[ht]
\centering
\includegraphics[width=\linewidth]{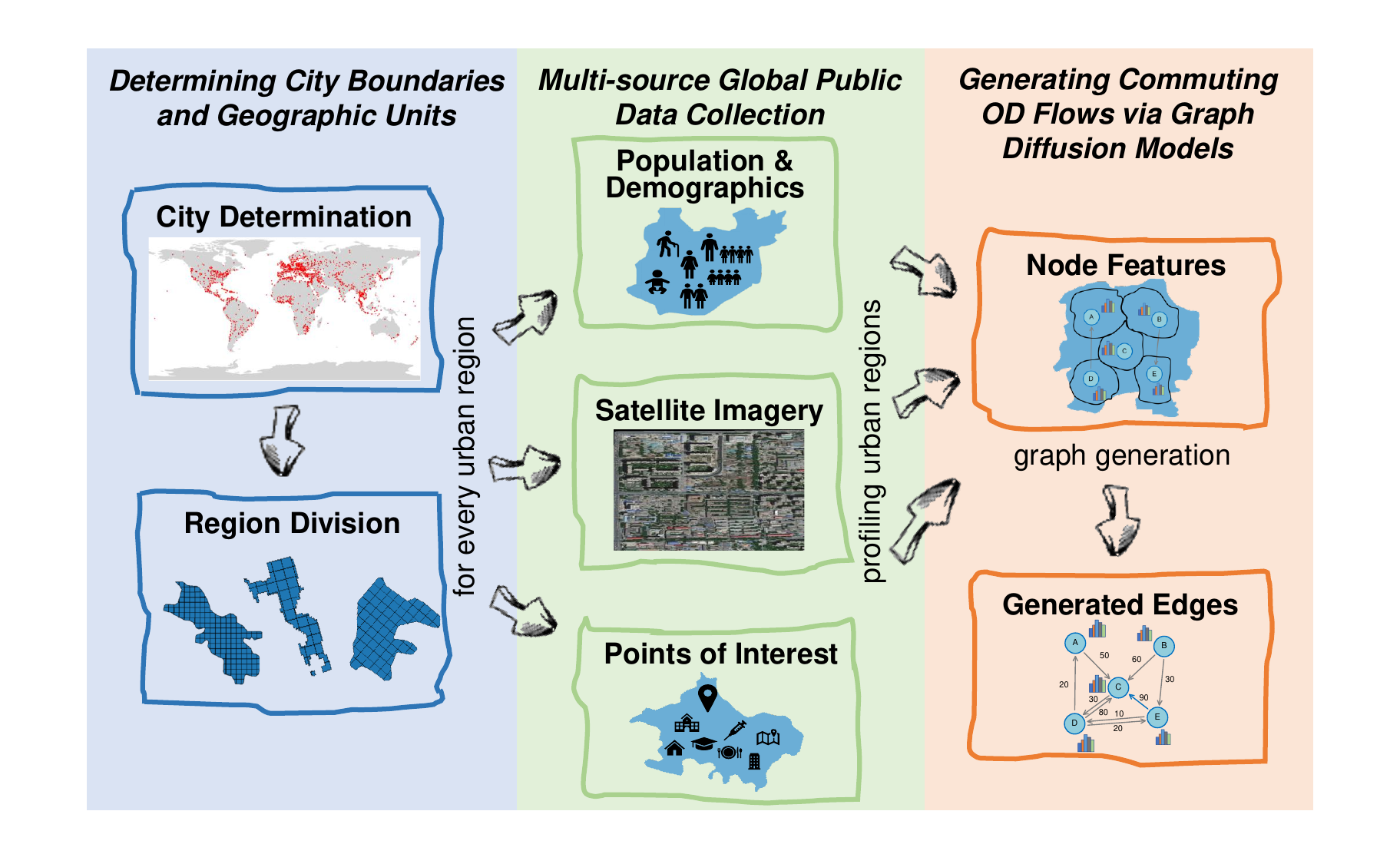}
\caption{An overview of the dataset construction pipeline.}
\label{fig:pipeline}
\end{figure}

\begin{figure}[ht]
\centering
\includegraphics[width=\linewidth]{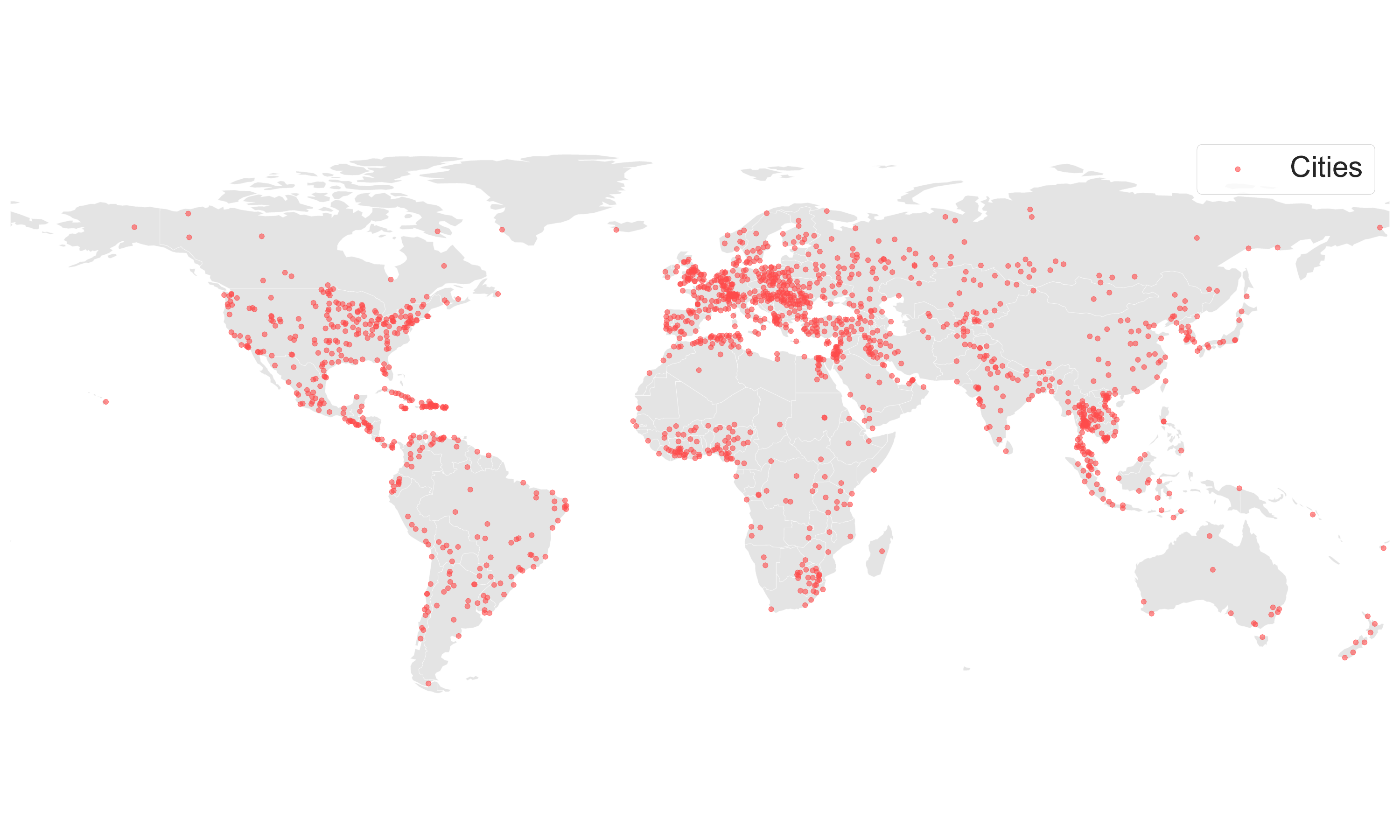}
\caption{An overview of the globally distributed cities included in the dataset.}
\label{fig:global_cities}
\end{figure}

\begin{figure}[ht]
\centering
\includegraphics[width=\linewidth]{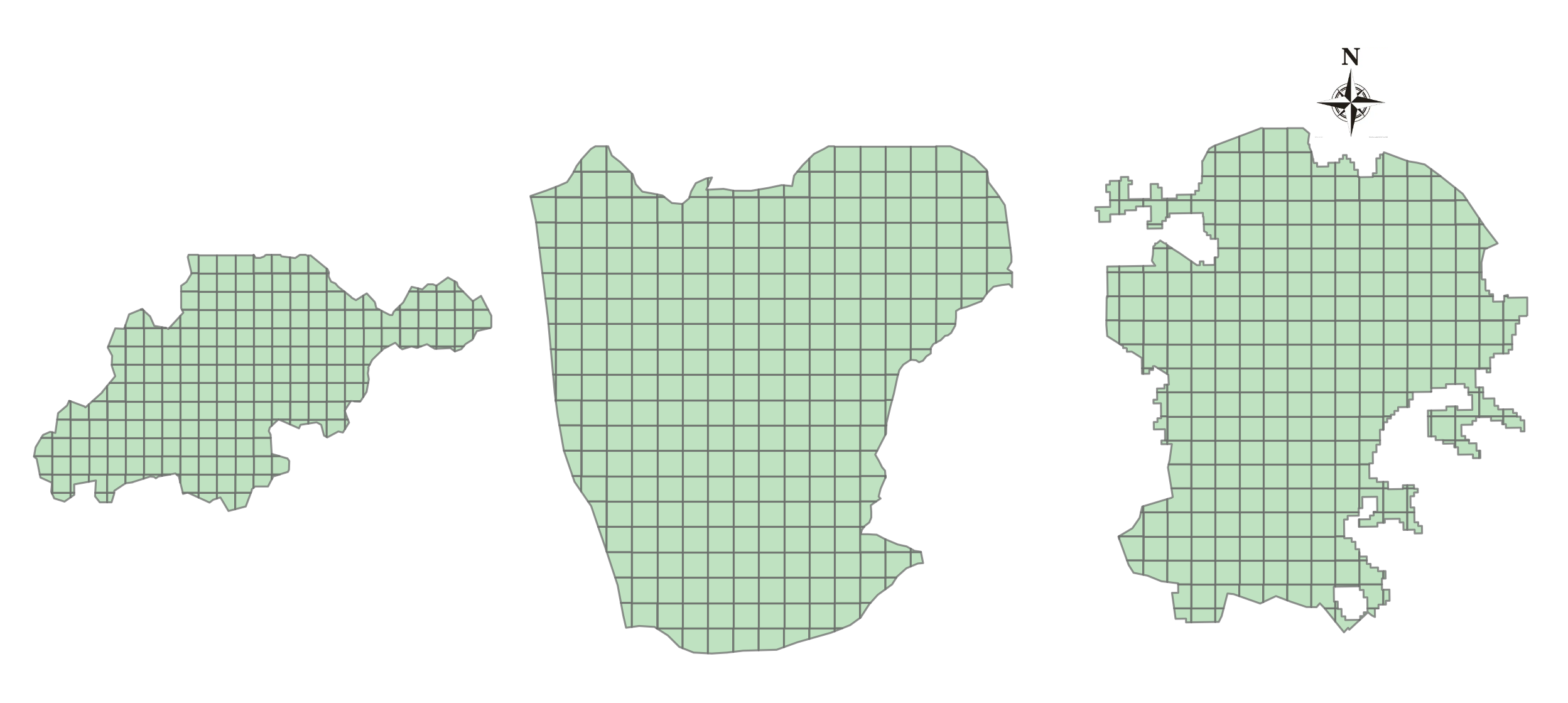}
\caption{Examples of the region division for cities.}
\label{fig:region_division_cases}
\end{figure}

\begin{figure}[ht]
    \centering
    \includegraphics[width=0.8\linewidth]{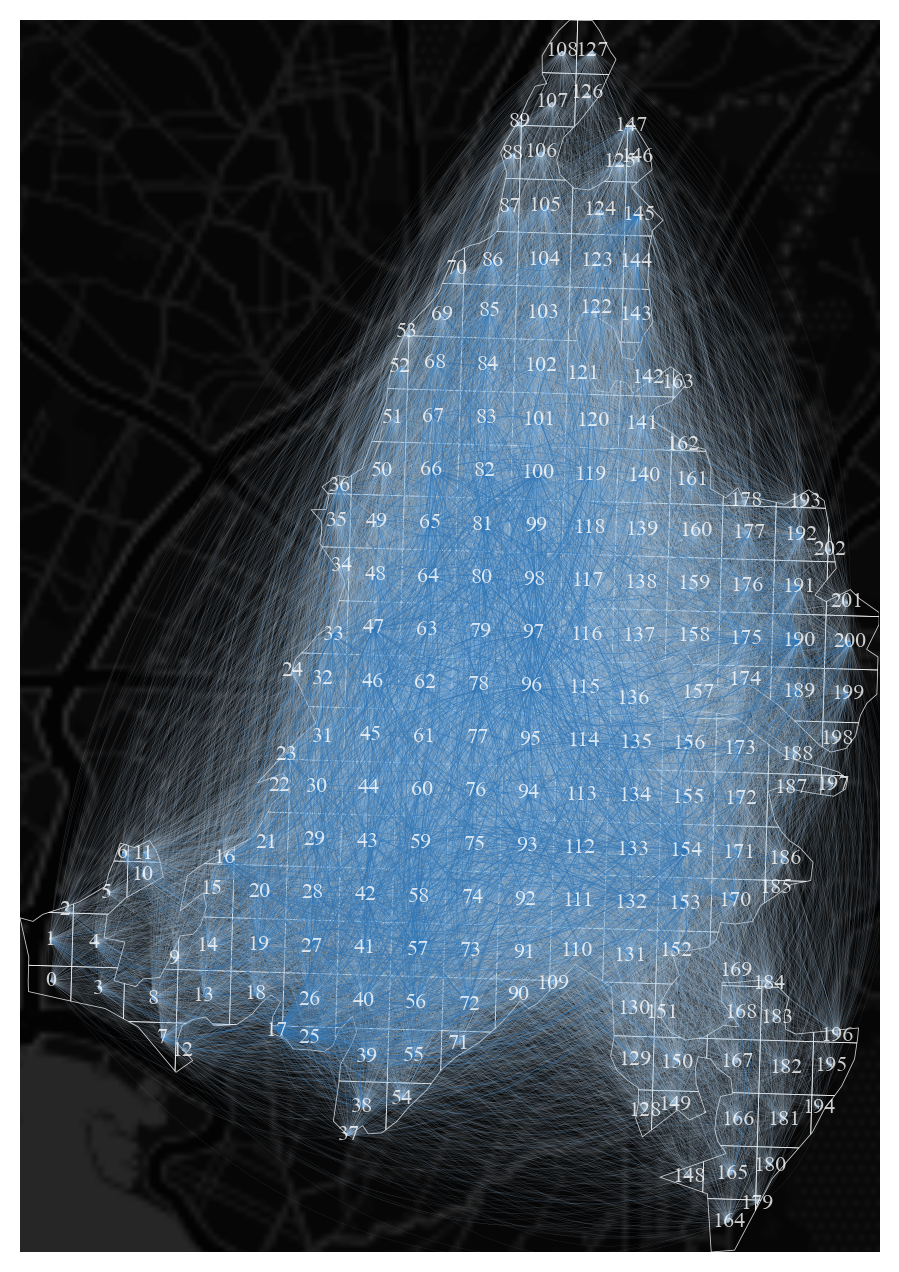}
    \caption{An visualization example of commuting OD flows in the city of Athens, Greece.}
    \label{fig:od_flow_vis_example}
\end{figure}

\begin{figure}[ht]
    \centering
    \begin{subfigure}[b]{0.265\textwidth}
        \centering
        \includegraphics[width=\linewidth]{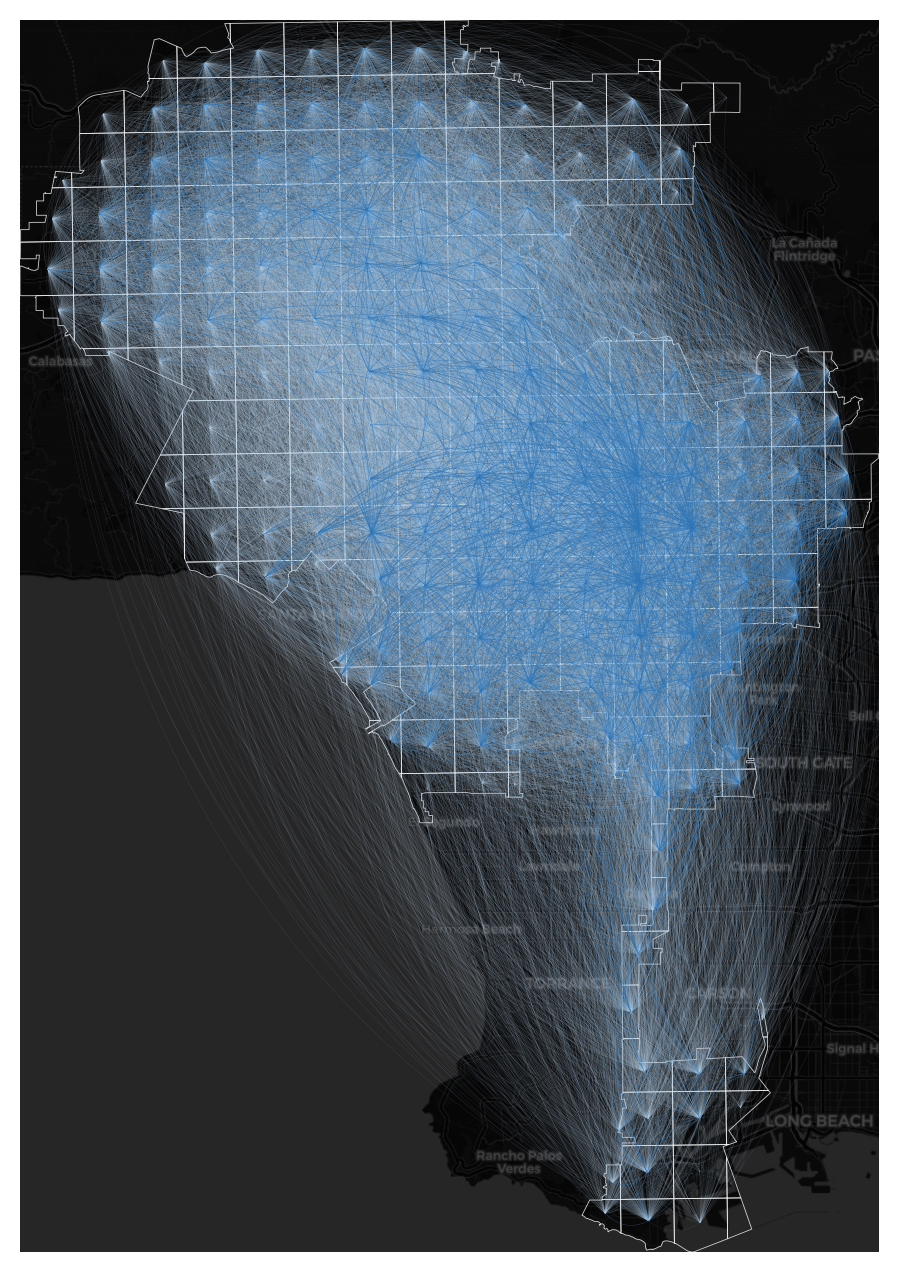}
        \caption{Los Angeles}
    \end{subfigure}
    \begin{subfigure}[b]{0.216\textwidth}
        \centering
        \includegraphics[width=\linewidth]{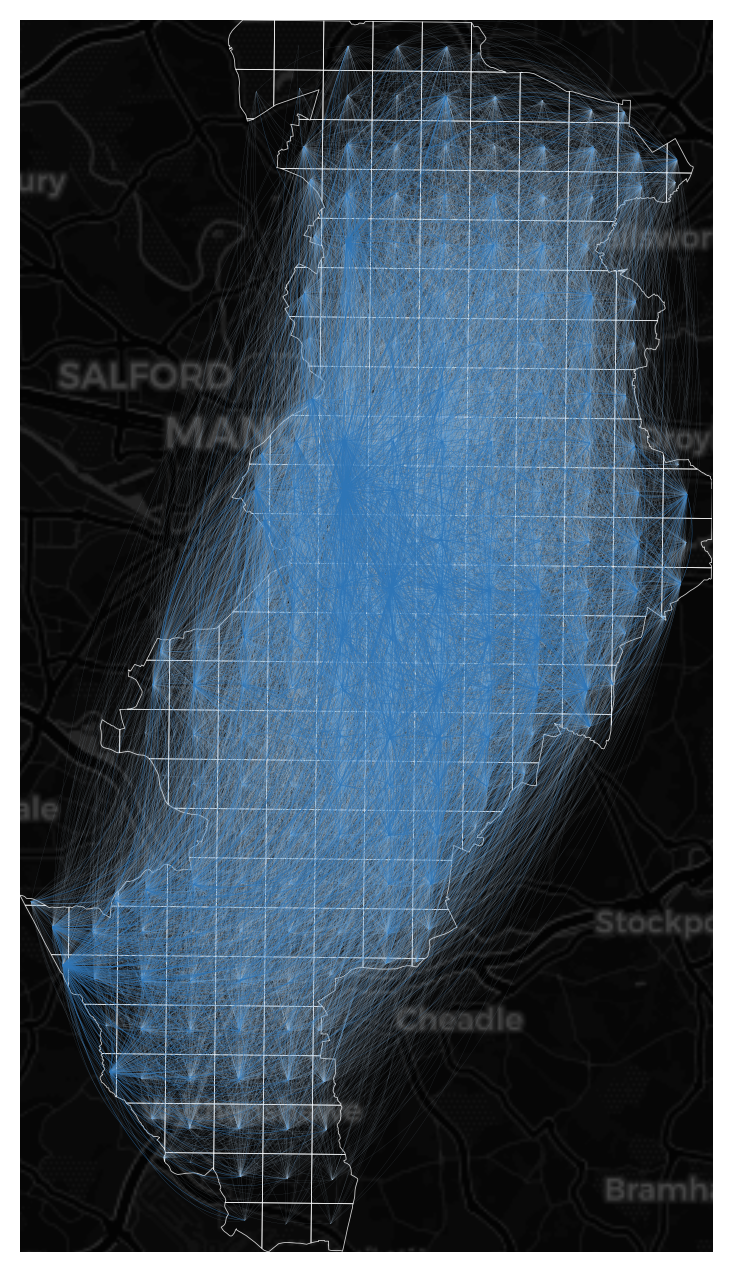}
        \caption{Manchester}
    \end{subfigure}
    \begin{subfigure}[b]{0.4\textwidth}
        \centering
        \includegraphics[width=\linewidth]{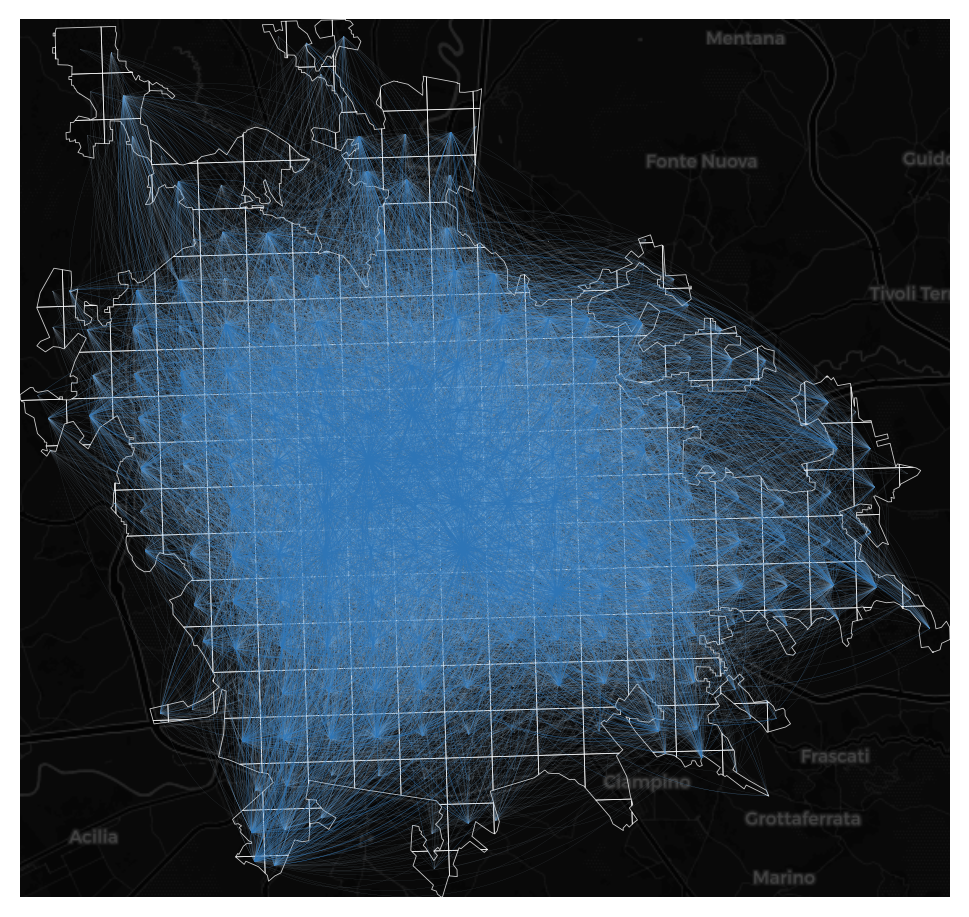}
        \caption{Rome}
    \end{subfigure}
    \caption{Examples of commuting OD flows in three representative cities.}
    \label{fig:od_flow_3example}
\end{figure}

\begin{figure}[ht]
\centering
\begin{subfigure}[b]{0.48\textwidth}
    \centering
    \includegraphics[width=\linewidth]{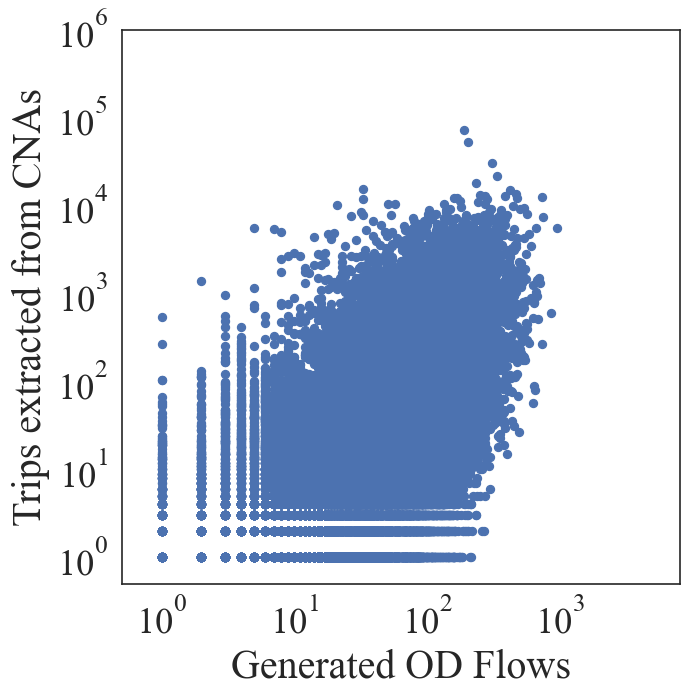}
    \caption{Beijing}
\end{subfigure}
\begin{subfigure}[b]{0.48\textwidth}
    \centering
    \includegraphics[width=\linewidth]{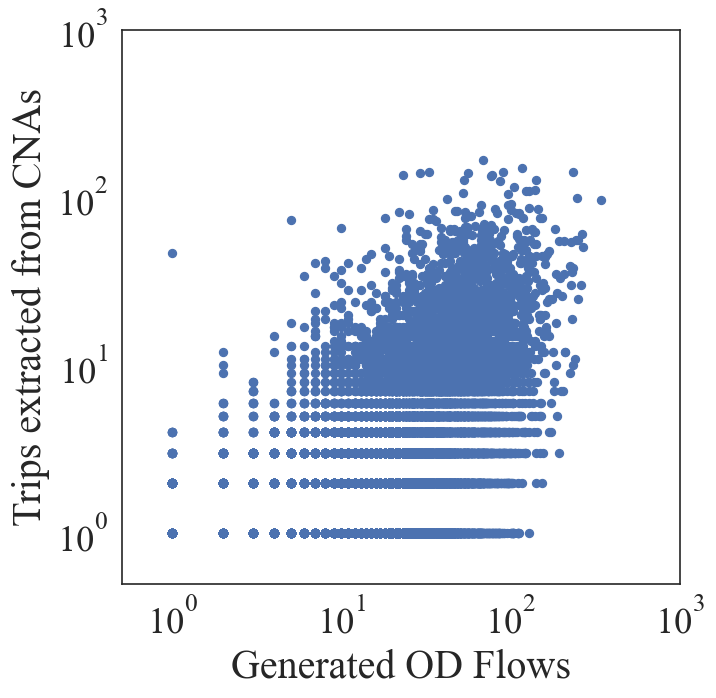}
    \caption{Shanghai}
\end{subfigure}
\begin{subfigure}[b]{0.48\textwidth}
    \centering
    \includegraphics[width=\linewidth]{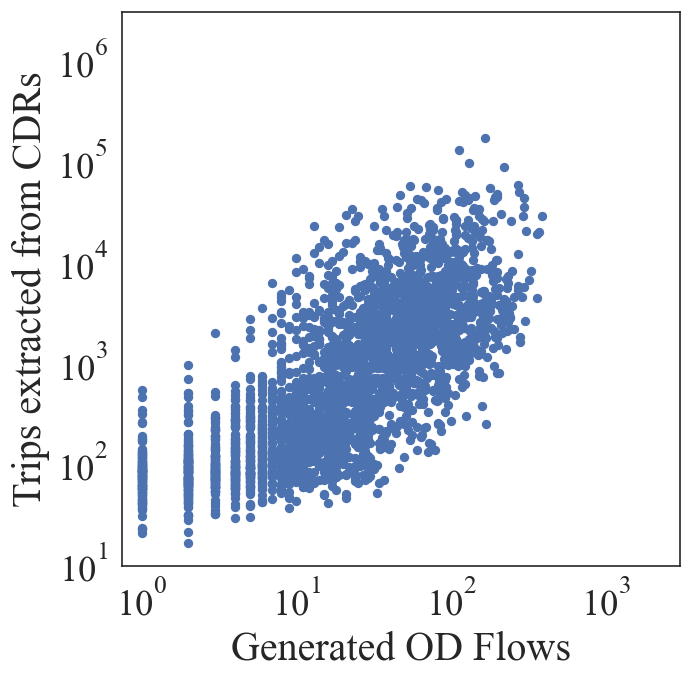}
    \caption{Rio de Janeiro}
\end{subfigure}
\begin{subfigure}[b]{0.48\textwidth}
    \centering
    \includegraphics[width=\linewidth]{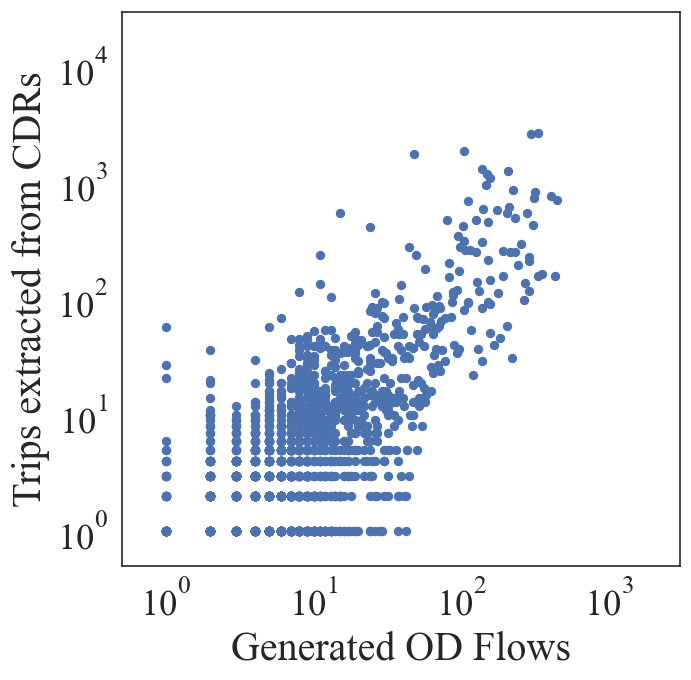}
    \caption{Senegal}
\end{subfigure}
\caption{Comparison between generated commuting OD flows and mobile location data.}
\label{fig:regression_analysis}
\end{figure}

\begin{figure}[ht]
\centering
\includegraphics[width=0.48\textwidth]{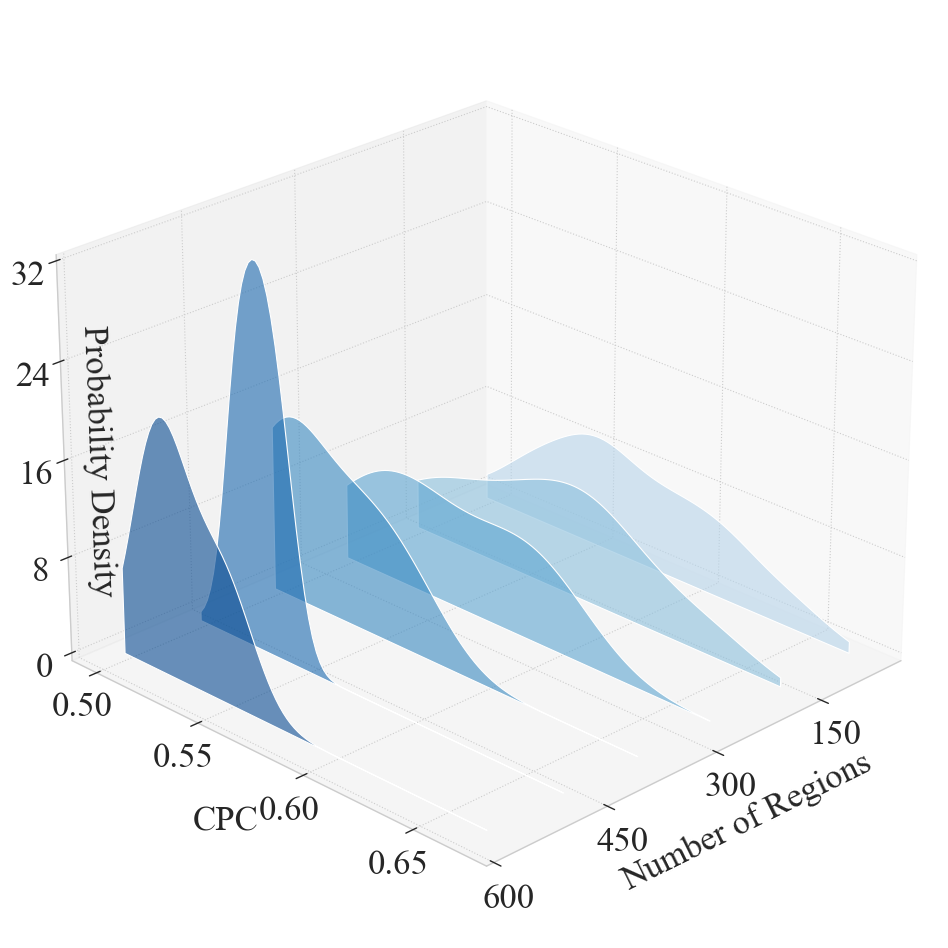}
\caption{CPC distribution against cities with different numbers of urban regions.}
\label{fig:cpc_distribution_size}
\end{figure}

\begin{figure}[ht]
\centering
\begin{subfigure}[b]{0.4\textwidth}
    \centering
    \includegraphics[width=\linewidth]{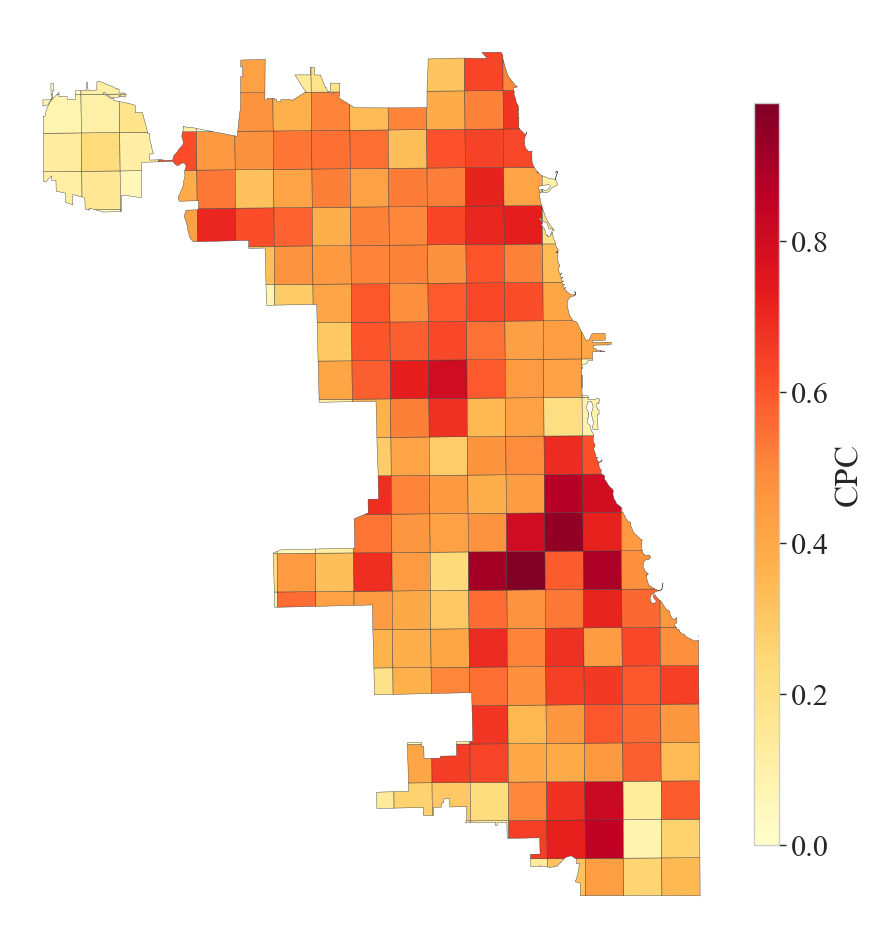}
    \caption{Chicago}
\end{subfigure}
\begin{subfigure}[b]{0.56\textwidth}
    \centering
    \includegraphics[width=\linewidth]{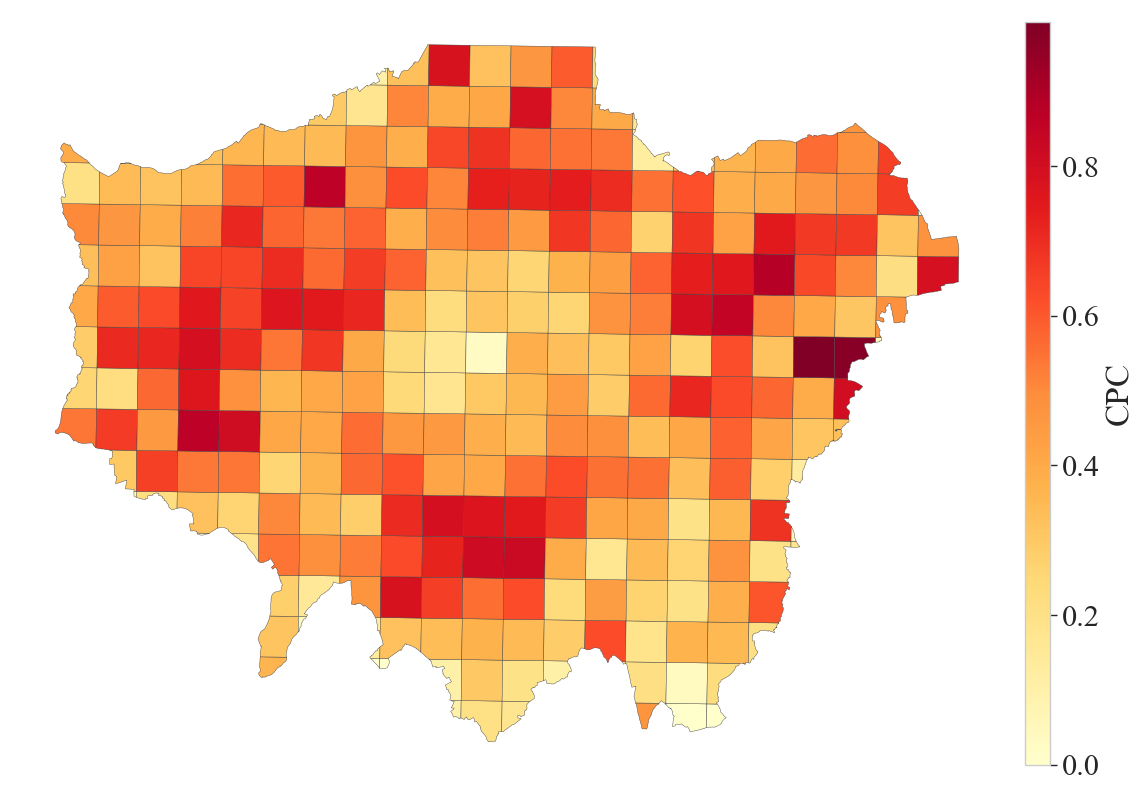}
    \caption{London}
\end{subfigure}
\caption{Spatial distribution of CPC in two representative cities.}
\label{fig:cpc_distribution_spatial}
\end{figure}

\begin{figure}[ht]
\centering
\begin{subfigure}[b]{0.48\textwidth}
    \centering
    \includegraphics[width=\linewidth]{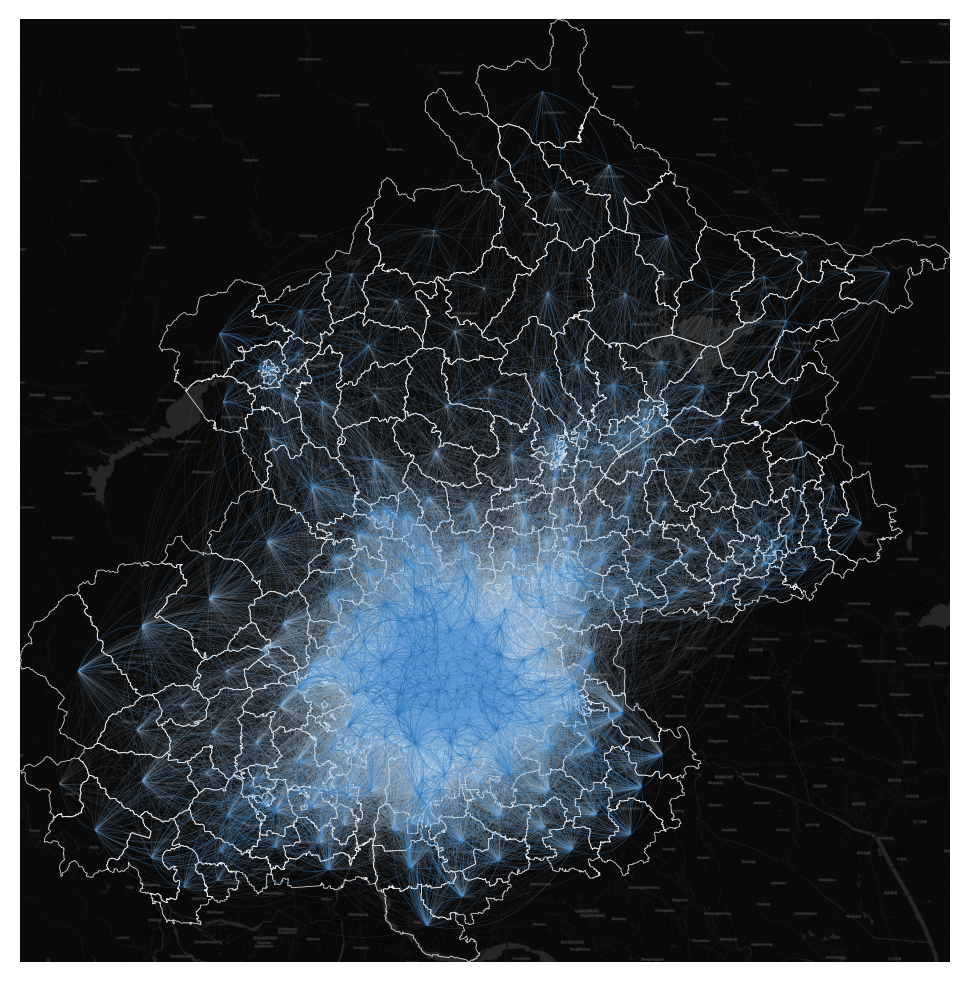}
    \caption{Trips extracted from cellular network access data.}
\end{subfigure}
\begin{subfigure}[b]{0.48\textwidth}
    \centering
    \includegraphics[width=\linewidth]{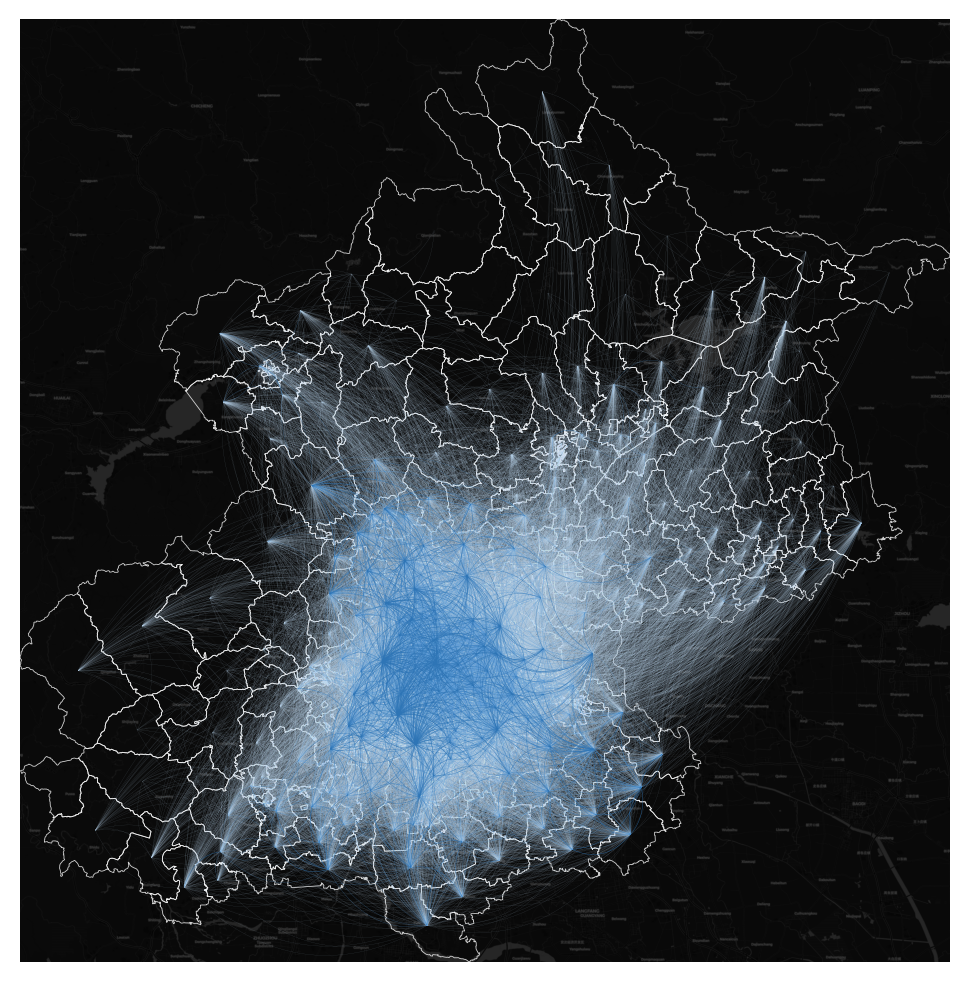}
    \caption{Generated commuting OD flows.}
\end{subfigure}
\caption{Geospatial visualization comparison between generated commuting OD flows and mobile location data for Beijing.}
\label{fig:geovisualization_beijing}
\end{figure}

\begin{figure}[ht]
    \centering
    \begin{subfigure}[b]{0.48\textwidth}
        \centering
        \includegraphics[width=\linewidth]{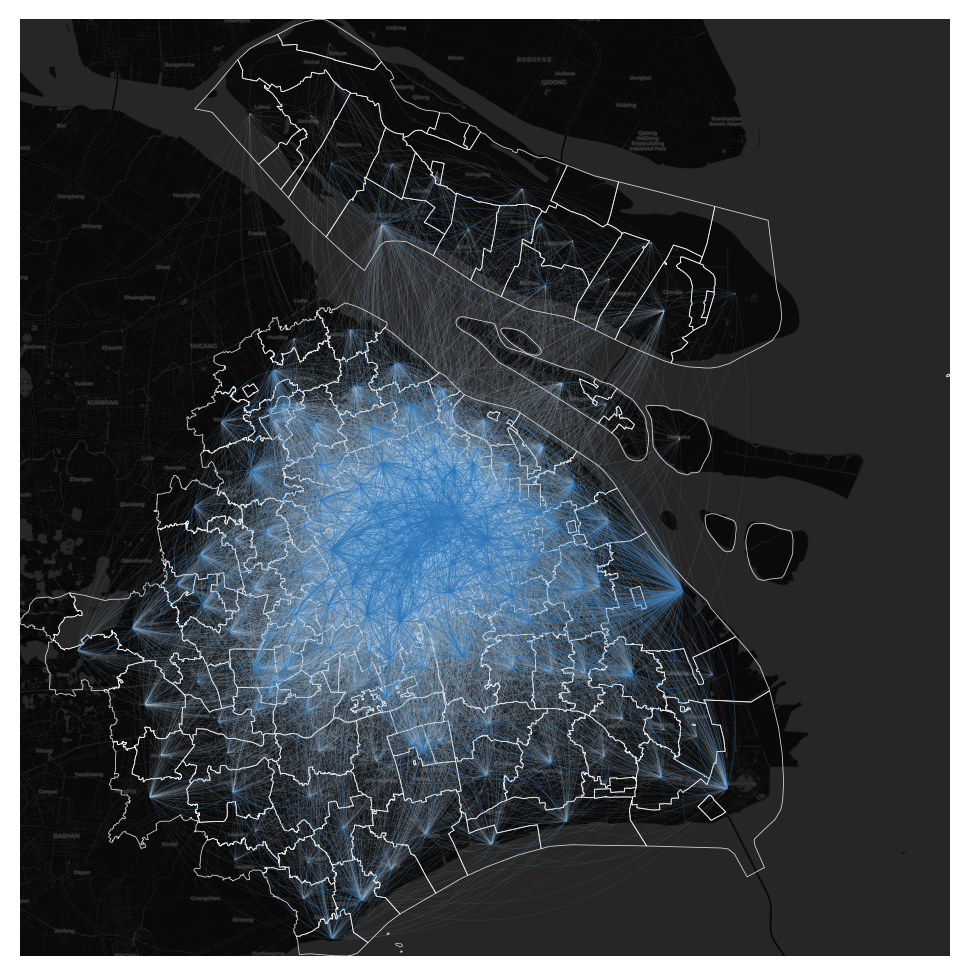}
        \caption{Trips extracted from cellular network access data.}
    \end{subfigure}
    \begin{subfigure}[b]{0.48\textwidth}
        \centering
        \includegraphics[width=\linewidth]{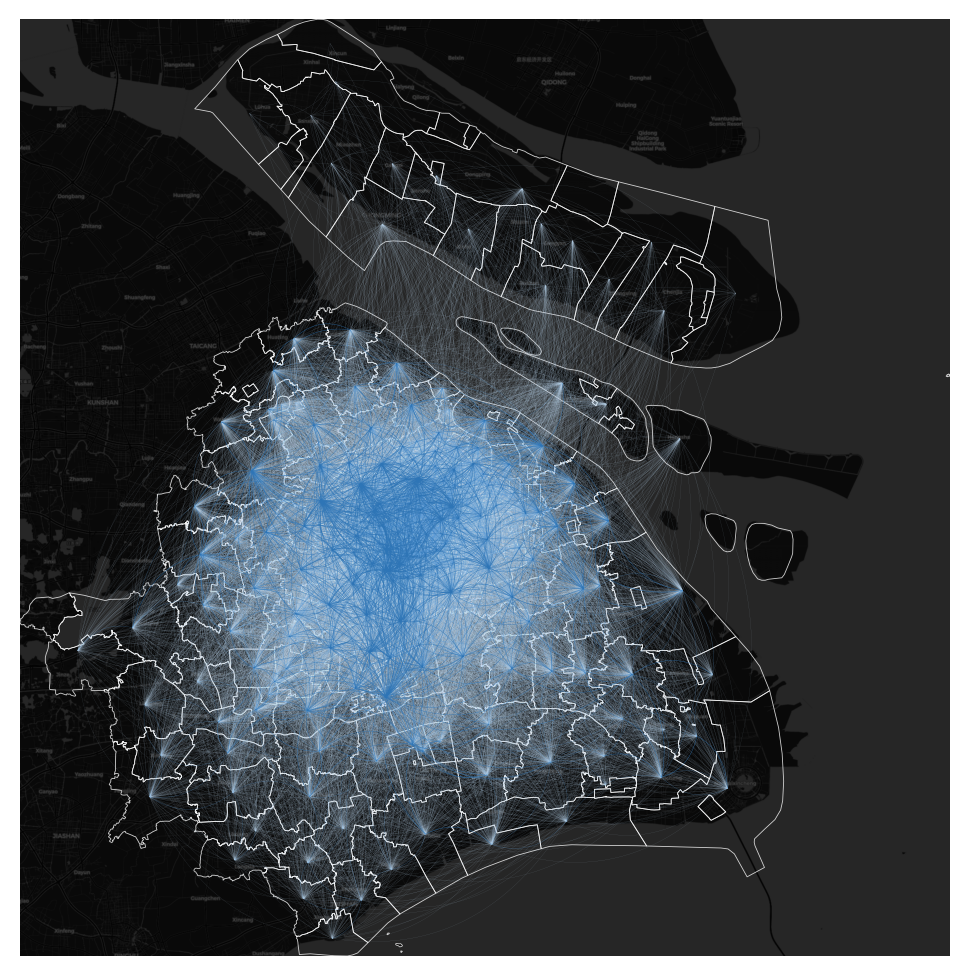}
        \caption{Generated commuting OD flows.}
    \end{subfigure}
    \caption{Geospatial visualization comparison between generated commuting OD flows and mobile location data for Shanghai.}
    \label{fig:geovisualization_shanghai}
\end{figure}

\begin{figure}[ht]
    \centering
    \begin{subfigure}[b]{0.48\textwidth}
        \centering
        \includegraphics[width=\linewidth]{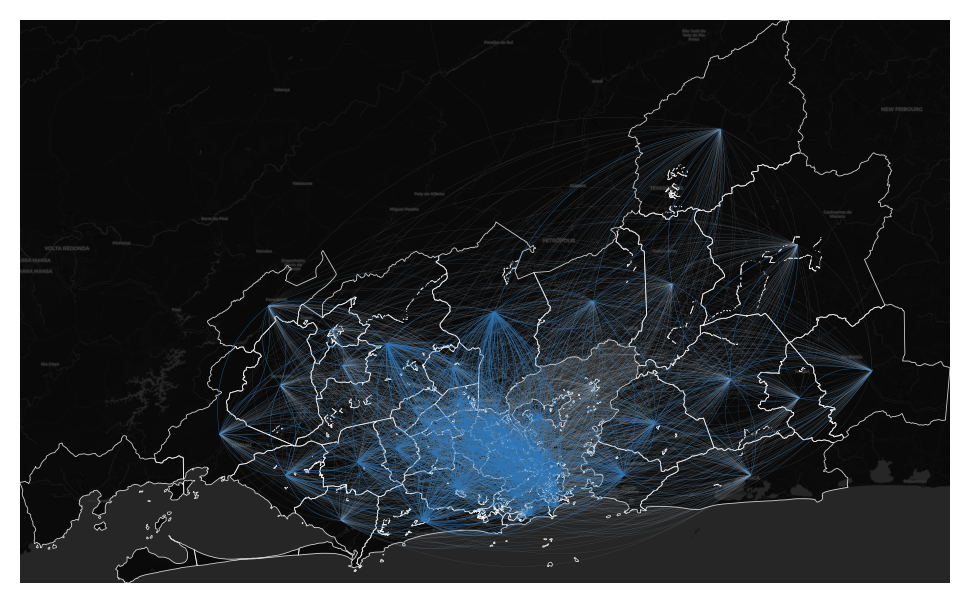}
        \caption{Commuting trips extracted from call detail records.}
    \end{subfigure}
    \begin{subfigure}[b]{0.48\textwidth}
        \centering
        \includegraphics[width=\linewidth]{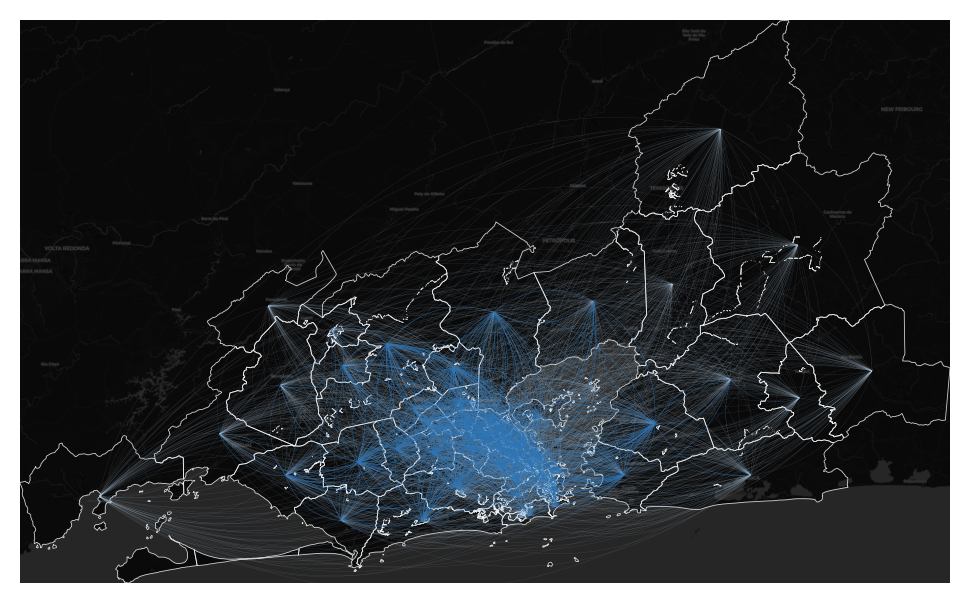}
        \caption{Generated commuting OD flows.}
    \end{subfigure}
    \caption{Geospatial visualization comparison between generated commuting OD flows and mobile location data for Rio de Janeiro.}
    \label{fig:geovisualization_rio}
\end{figure}

\begin{figure}[ht]
    \centering
    \begin{subfigure}[b]{0.48\textwidth}
        \centering
        \includegraphics[width=\linewidth]{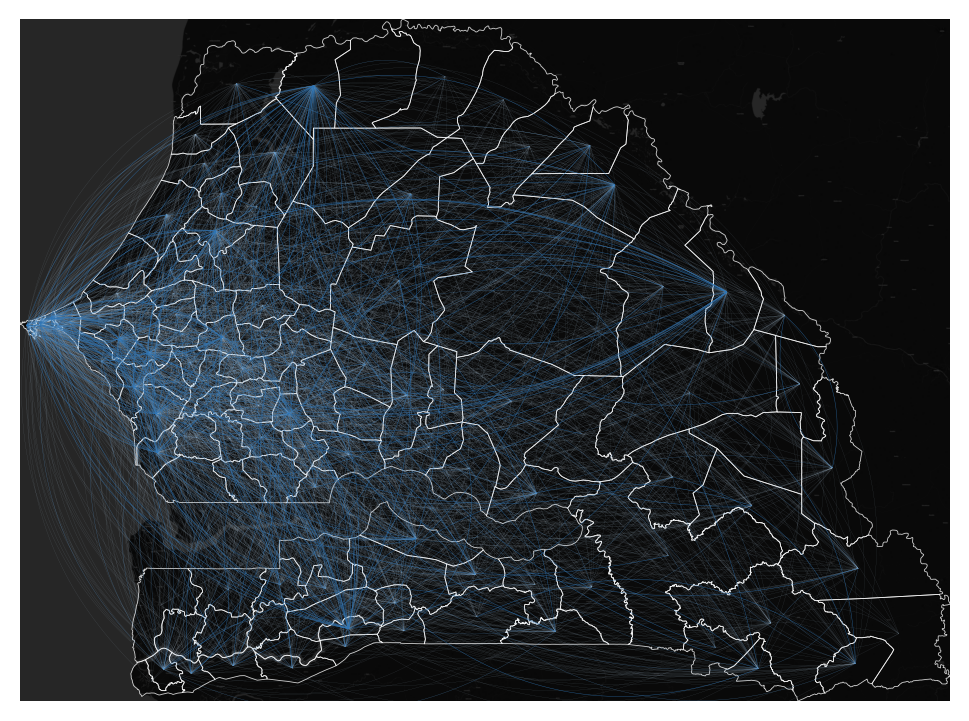}
        \caption{Commuting trips extracted from call detail records.}
    \end{subfigure}
    \begin{subfigure}[b]{0.48\textwidth}
        \centering
        \includegraphics[width=\linewidth]{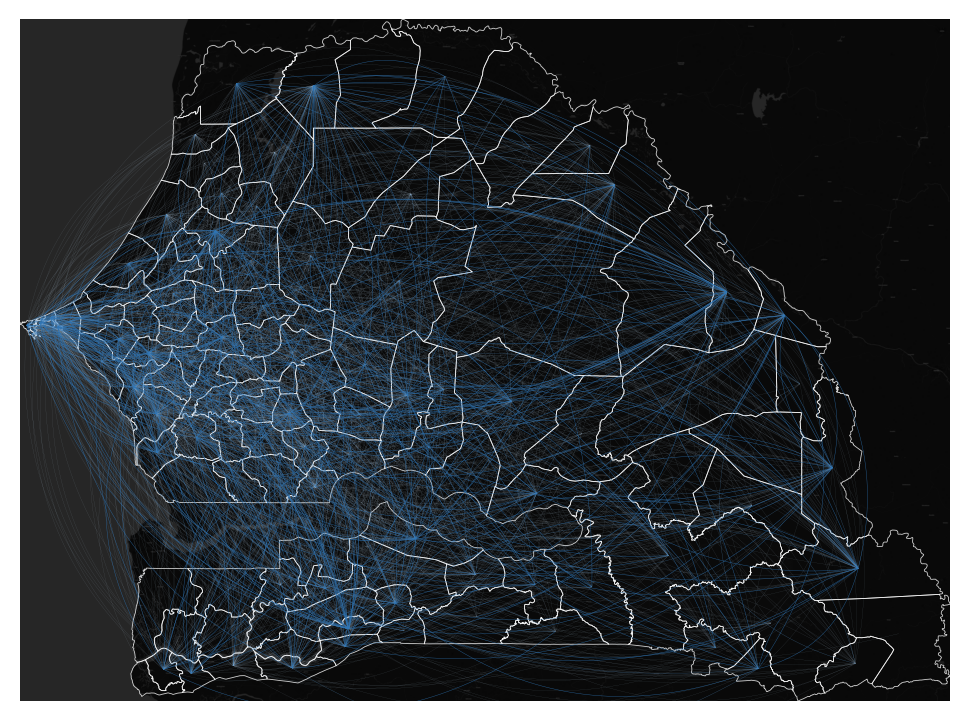}
        \caption{Generated commuting OD flows.}
    \end{subfigure}
    \caption{Geospatial visualization comparison between generated commuting OD flows and mobile location data for Senegal.}
    \label{fig:geovisualization_senegal}
\end{figure}

\begin{figure}[ht]
    \centering
    \begin{subfigure}[b]{0.24\textwidth}
        \centering
        \includegraphics[height=4cm]{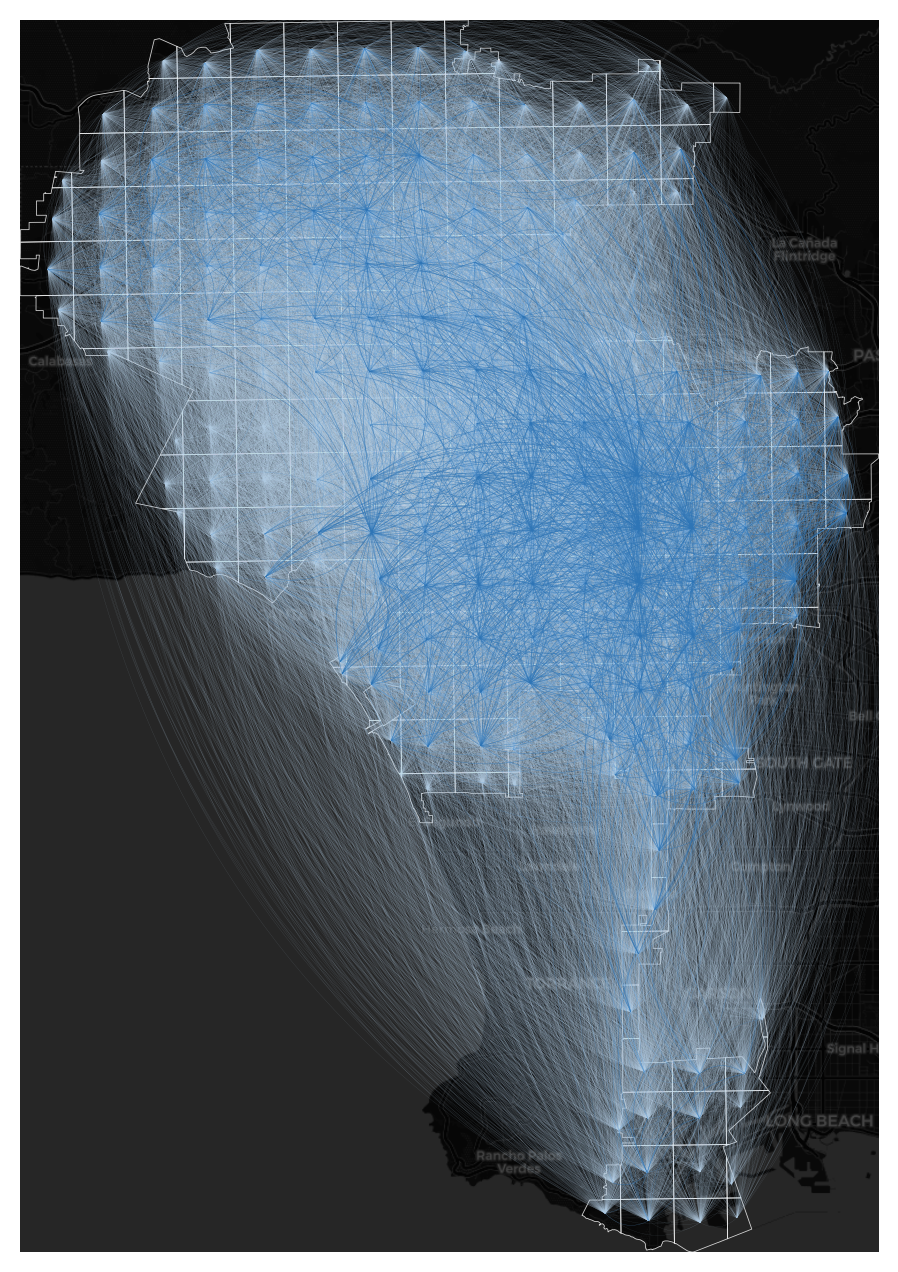}
        \caption{Los Angeles}
    \end{subfigure}
    \begin{subfigure}[b]{0.24\textwidth}
        \centering
        \includegraphics[height=4cm]{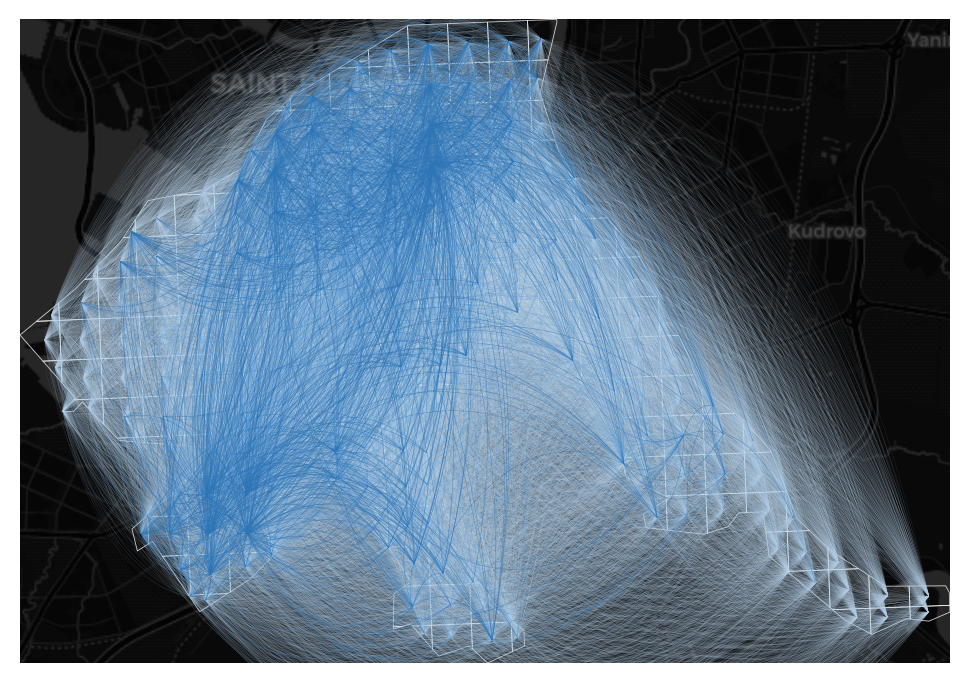}
        \caption{St. Petersburg}
    \end{subfigure}
    \begin{subfigure}[b]{0.24\textwidth}
        \centering
        \includegraphics[width=3cm,height=4cm]{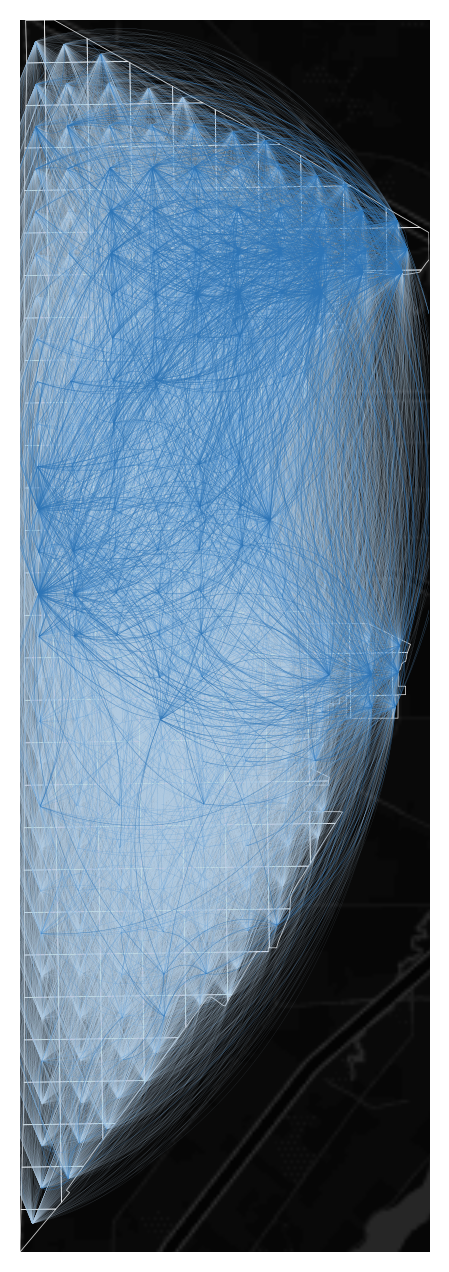}
        \caption{Hobart}
    \end{subfigure}
    \begin{subfigure}[b]{0.24\textwidth}
        \centering
        \includegraphics[height=4cm]{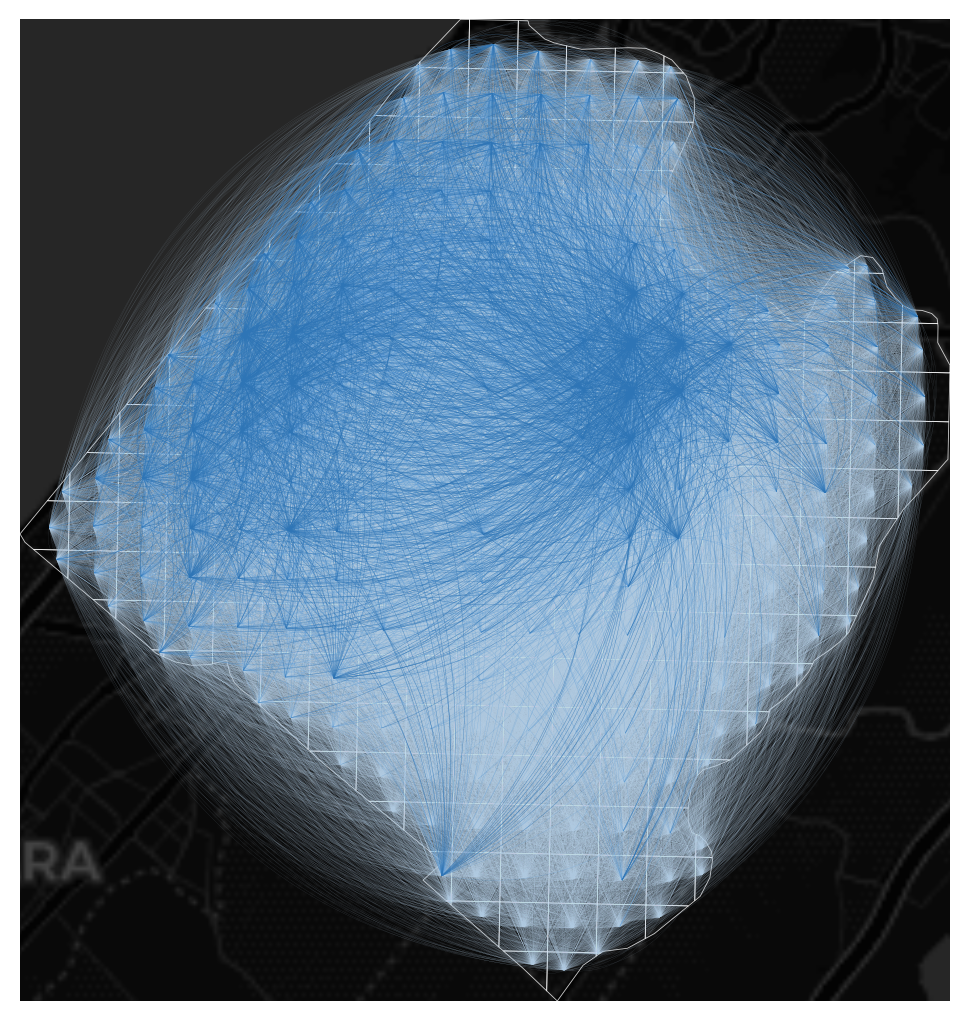}
        \caption{Rabat}
    \end{subfigure}
    \caption{Geospatial visualization of commuting OD flows of representative polycentric cities.}
    \label{fig:geovis_polycentric_cities}
\end{figure}

\begin{figure}[ht]
    \centering
    \begin{subfigure}[b]{0.48\textwidth}
        \centering
        \includegraphics[width=\linewidth]{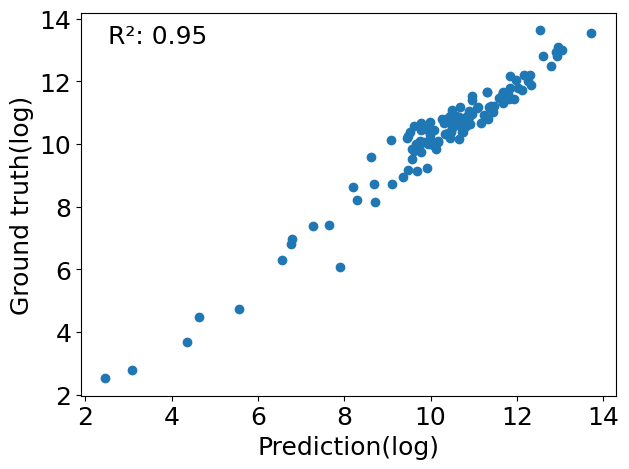}
        \caption{Population distribution prediction during work hours for Chengdu.}
    \end{subfigure}
    \begin{subfigure}[b]{0.48\textwidth}
        \centering
        \includegraphics[width=\linewidth]{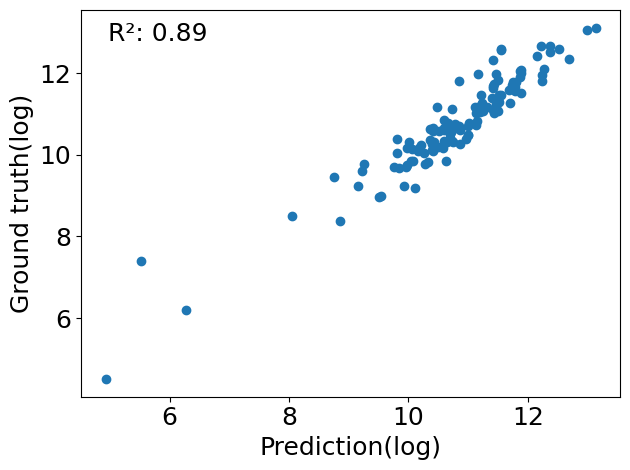}
        \caption{Population distribution prediction during work hours for Guangzhou.}
    \end{subfigure}
    
    \begin{subfigure}[b]{0.48\textwidth}
        \centering
        \includegraphics[width=\linewidth]{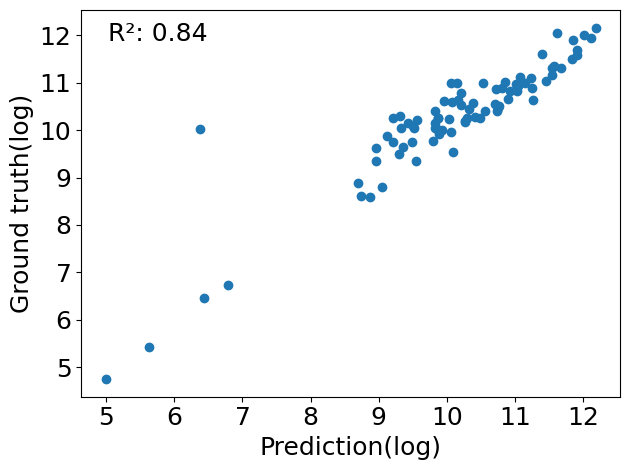}
        \caption{Population distribution prediction during work hours for Jinan.}
    \end{subfigure}
    \begin{subfigure}[b]{0.48\textwidth}
        \centering
        \includegraphics[width=\linewidth]{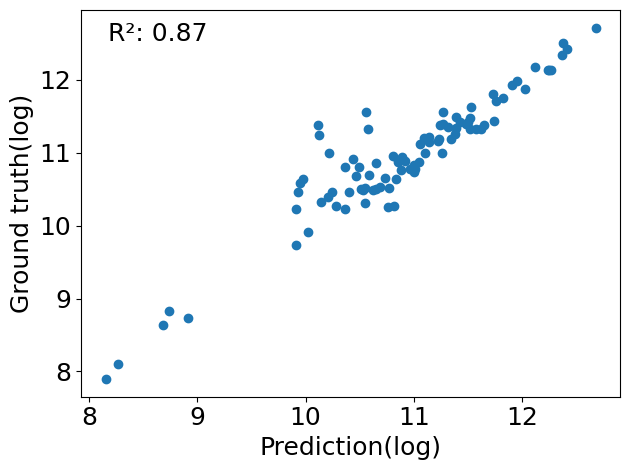}
        \caption{Population distribution prediction during work hours for Qingdao.}
    \end{subfigure}
    \caption{Downstream task results of population distribution prediction during work hours.}
    \label{fig:population_prediction}
\end{figure}

\begin{figure}[ht]
    \centering
    \begin{subfigure}[b]{0.32\textwidth}
        \centering
        \includegraphics[width=\linewidth]{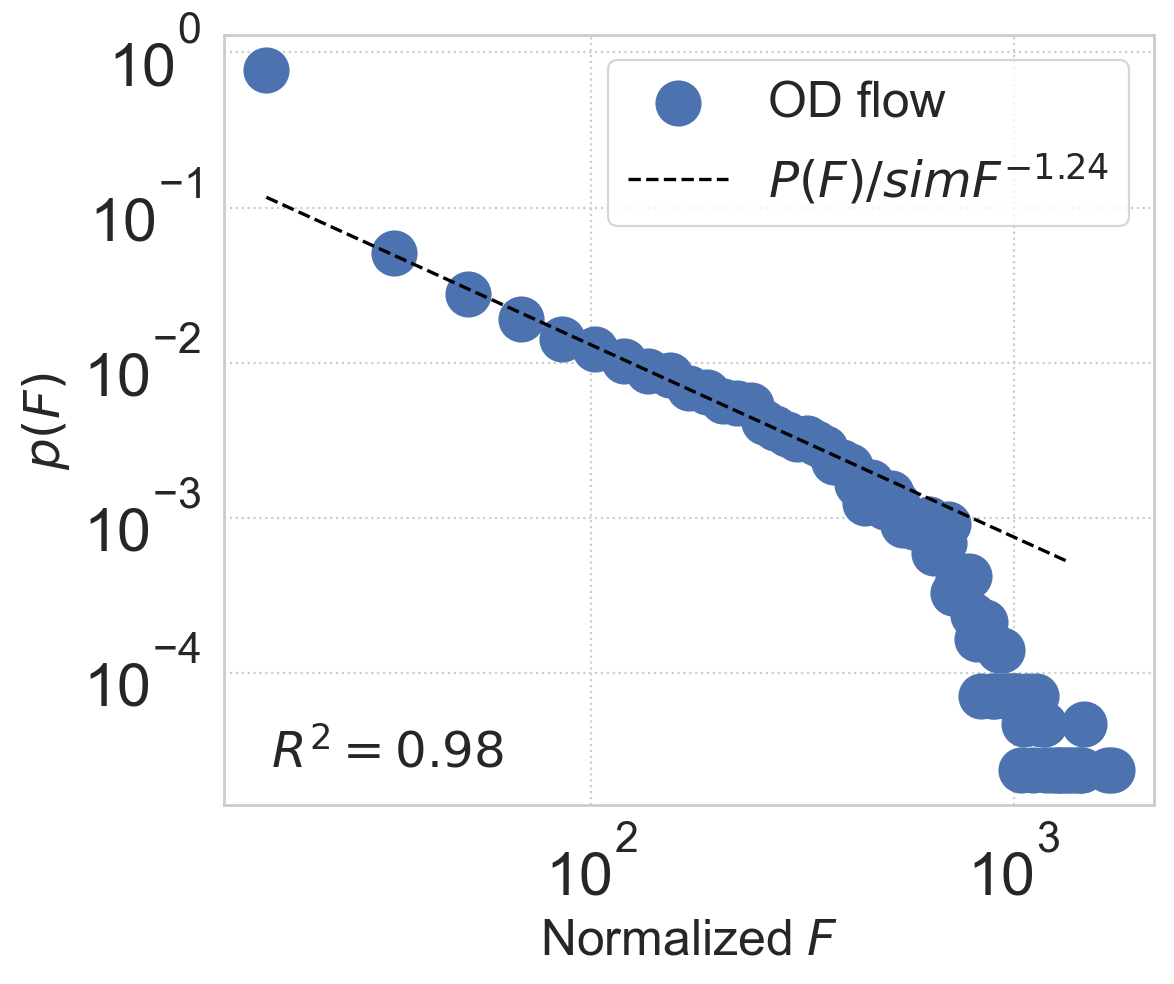}
        \caption{Chengdu}
    \end{subfigure}
    \begin{subfigure}[b]{0.32\textwidth}
        \centering
        \includegraphics[width=\linewidth]{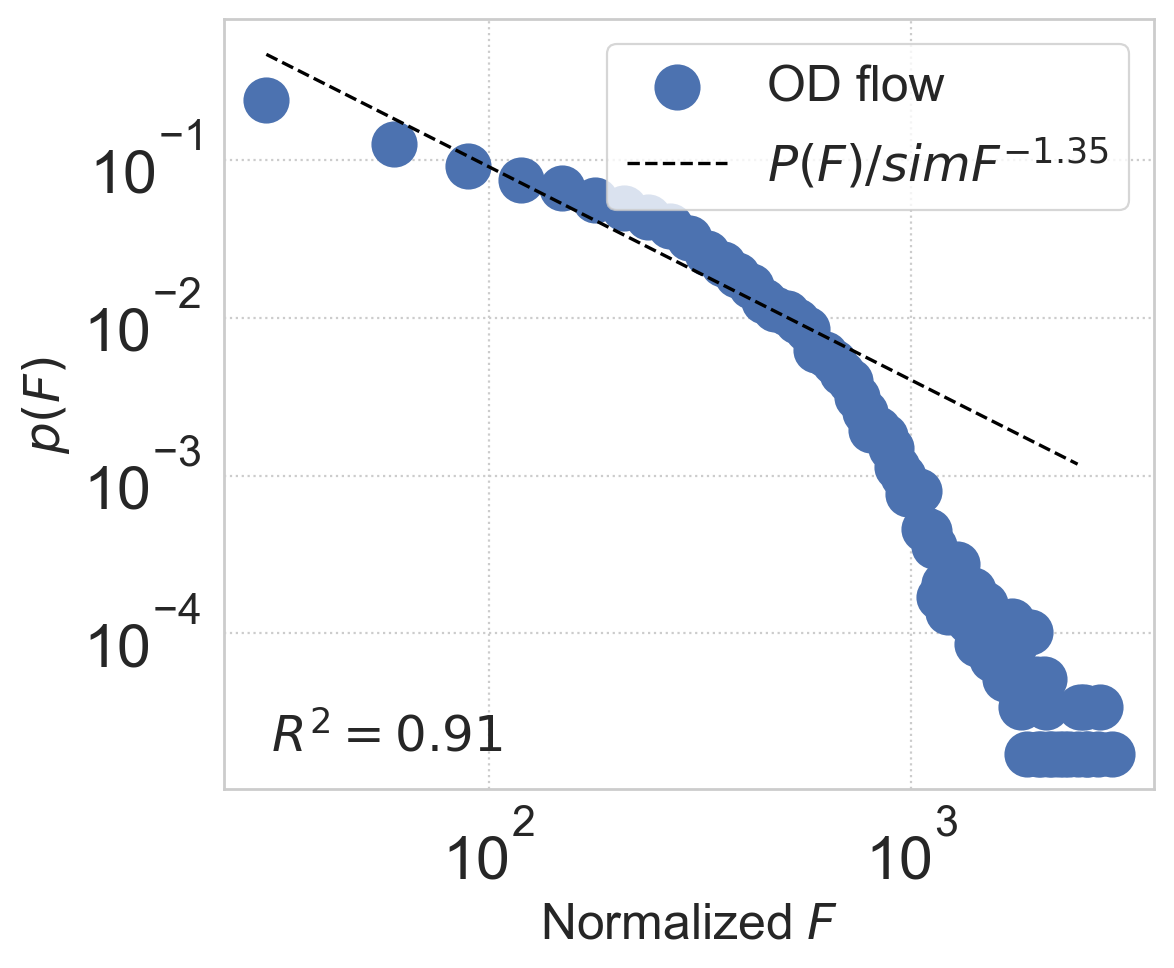}
        \caption{Guangzhou}
    \end{subfigure}
    \begin{subfigure}[b]{0.32\textwidth}
        \centering
        \includegraphics[width=\linewidth]{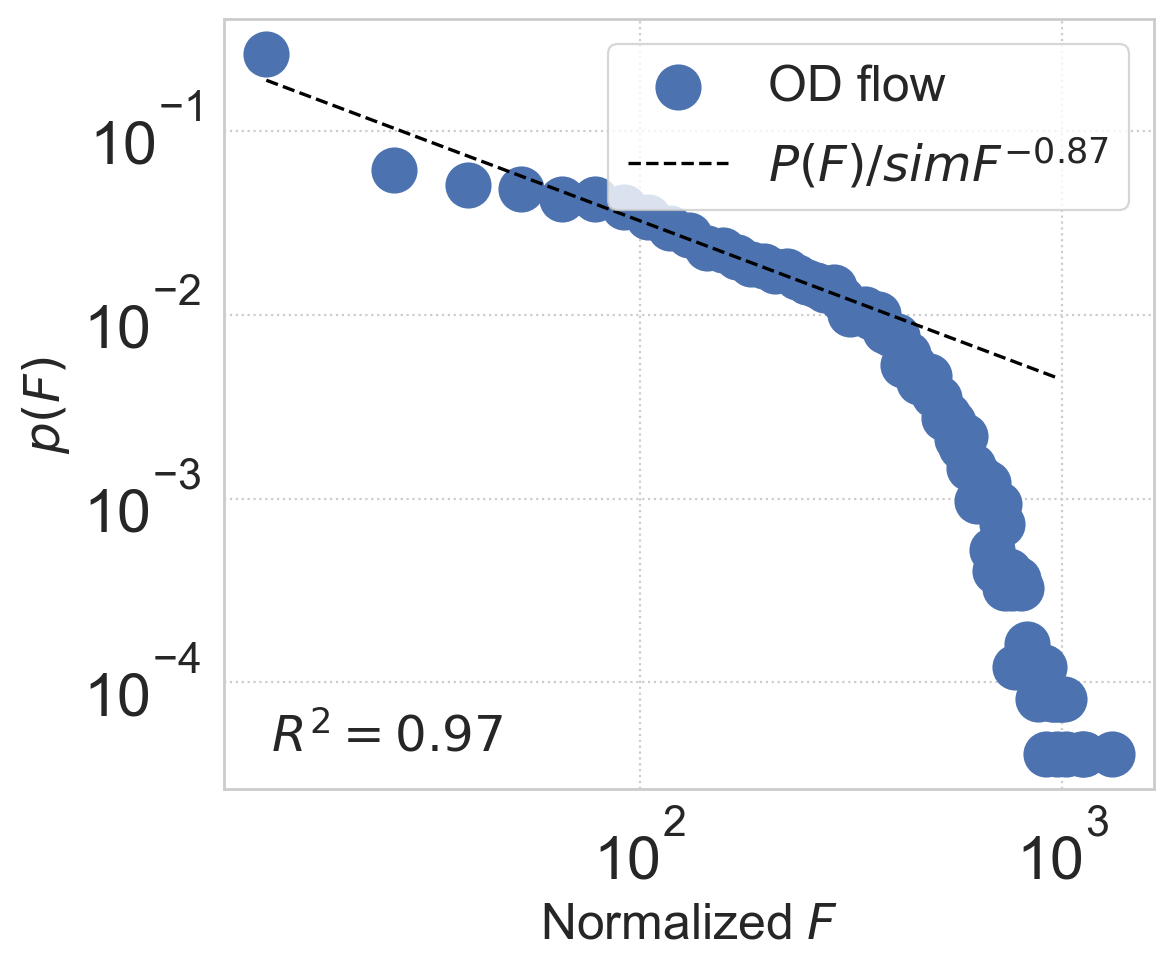}
        \caption{Qingdao}
    \end{subfigure}
    
    \begin{subfigure}[b]{0.32\textwidth}
        \centering
        \includegraphics[width=\linewidth]{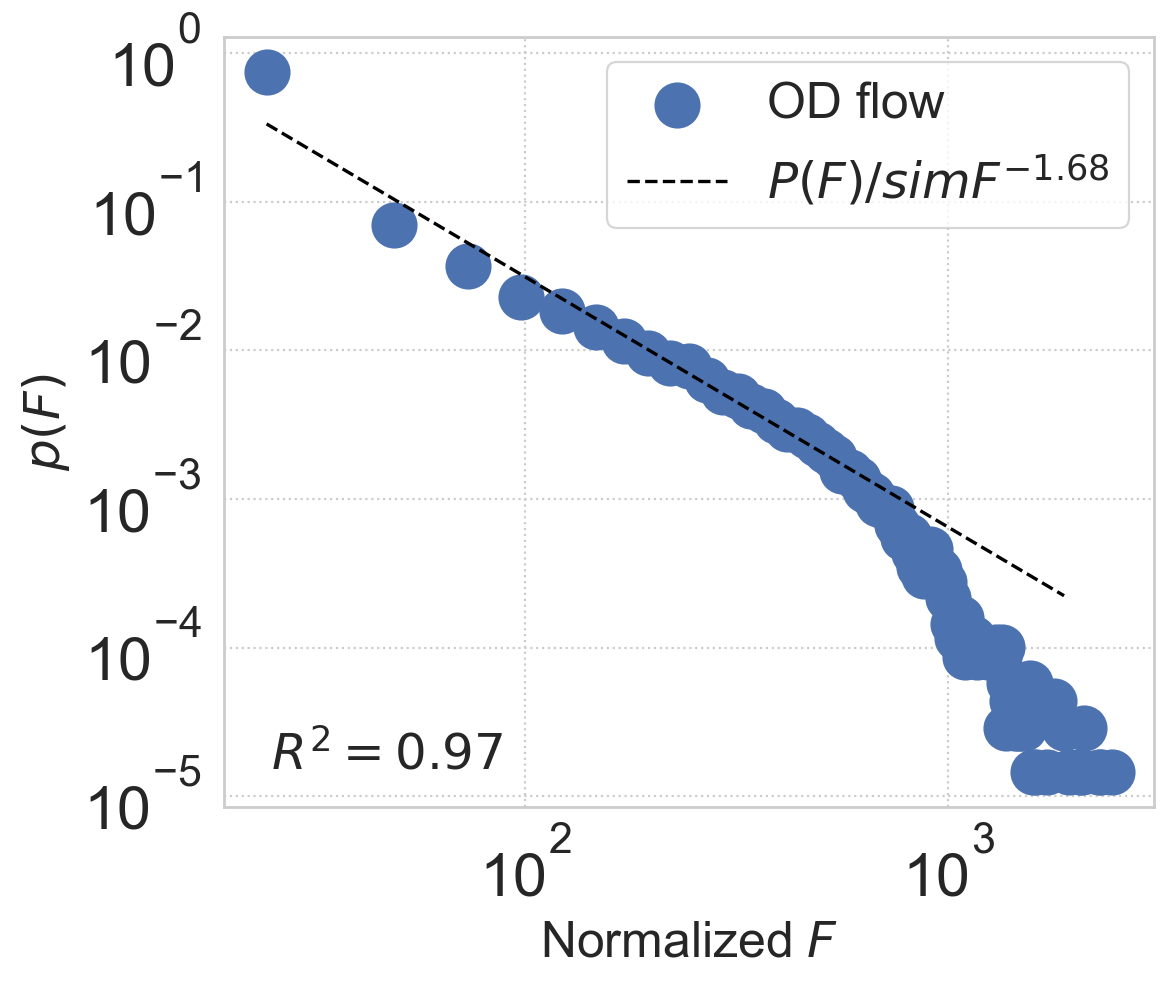}
        \caption{Hong Kong}
    \end{subfigure}
    \begin{subfigure}[b]{0.32\textwidth}
        \centering
        \includegraphics[width=\linewidth]{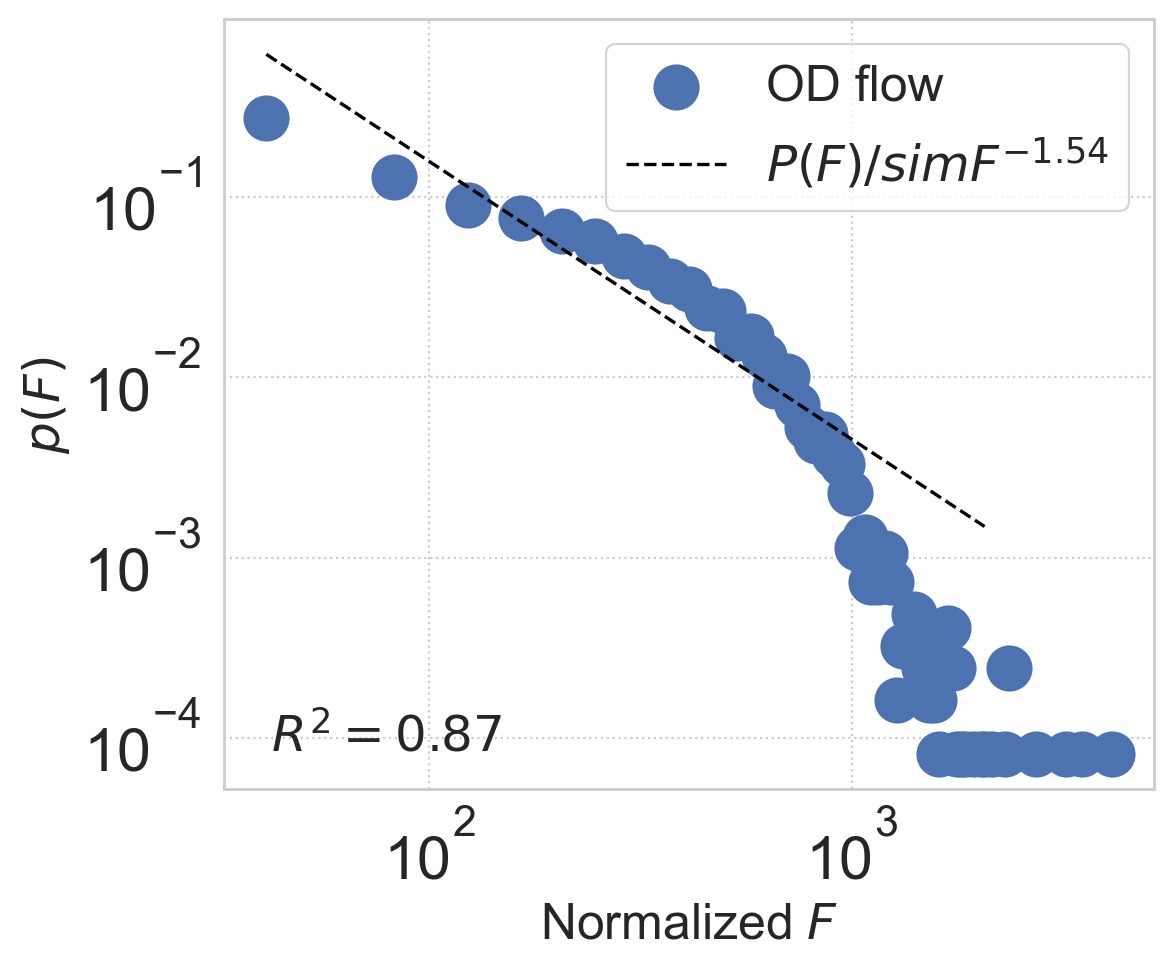}
        \caption{Tokyo}
    \end{subfigure}
    \begin{subfigure}[b]{0.32\textwidth}
        \centering
        \includegraphics[width=\linewidth]{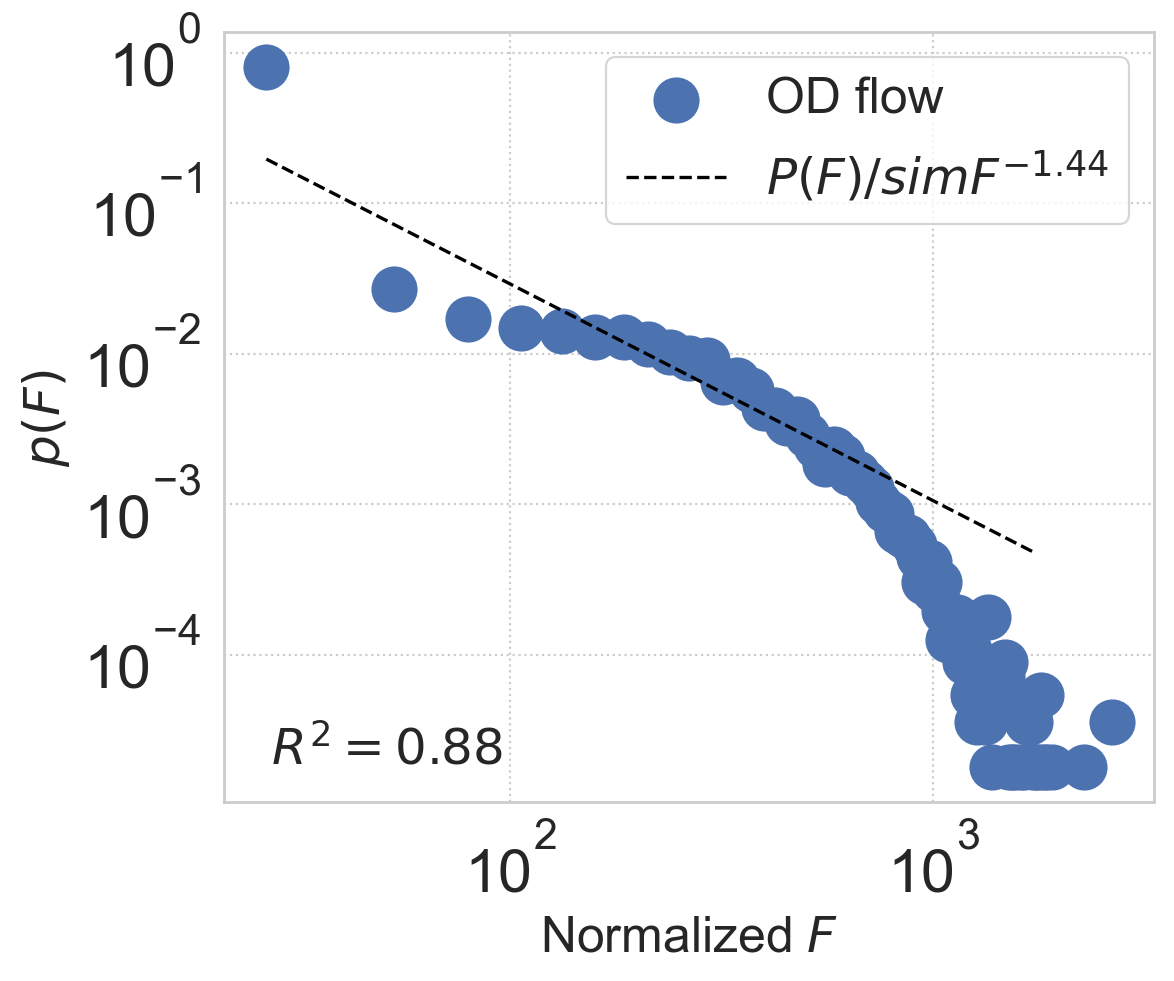}
        \caption{Singapore}
    \end{subfigure}
    \caption{Scale-free property analysis of generated commuting networks in different cities.}
    \label{fig:scaling_analysis}
\end{figure}

\begin{table}[ht]
\centering
\caption{An example of the city boundary and region division for the city of Athens, Greece.}
\label{tab:region_division_cases}
\begin{tabular}{|c|c|c|}
\toprule
index & locations & geometry \\
\midrule
0 & 0-4 & POLYGON ((23.6773 37.94935, 23.67718 37.94611,... \\
1 & 0-5 & POLYGON ((23.67748 37.95436, 23.6773 37.94935,... \\
2 & 0-6 & POLYGON ((23.67748 37.95436, 23.67454 37.95443... \\
3 & 1-4 & POLYGON ((23.68345 37.94446, 23.68188 37.94514... \\
4 & 1-5 & POLYGON ((23.6773 37.94935, 23.67748 37.95436,... \\
... & ... & ... \\
91 & 9-5 & POLYGON ((23.7279 37.94819, 23.72808 37.9532, ... \\
92 & 9-6 & POLYGON ((23.72808 37.9532, 23.72827 37.95821,... \\
93 & 9-7 & POLYGON ((23.72827 37.95821, 23.72846 37.96322... \\
94 & 9-8 & POLYGON ((23.72846 37.96322, 23.72864 37.96822... \\
95 & 9-9 & POLYGON ((23.72864 37.96822, 23.72883 37.97323... \\
\midrule
\multicolumn{3}{|l|}{
\begin{minipage}{0.63\linewidth}
<Geographic 2D CRS: EPSG:4326> \\
Name: WGS 84 \\
Axis Info [ellipsoidal]: \\
- Lat[north]: Geodetic latitude (degree) \\
- Lon[east]: Geodetic longitude (degree) \\
Area of Use: \\
- name: World. \\
- bounds: (-180.0, -90.0, 180.0, 90.0) \\
Datum: World Geodetic System 1984 ensemble \\
- Ellipsoid: WGS 84 \\
- Prime Meridian: Greenwich \\
\end{minipage}
}
\\
\bottomrule
\end{tabular}
\end{table}

\begin{table}[ht]
\centering
\caption{An example of the commuting OD flows in the city of Athens, Greece.}
\label{tab:od_flow_matrix}
\begin{tabular}{|c|c|c|c|c|c|c|c|}
\toprule
Row/Col & Col1 & Col2 & Col3 & ... & Col94 & Col95 & Col96\\
\midrule
Row1  & 0   & 112 & 139 & ... & 79  & 66  & 58 \\ \hline
Row2  & 113 & 0   & 142 & ... & 75  & 54  & 63 \\ \hline
Row3  & 136 & 171 & 0   & ... & 63  & 45  & 50 \\ \hline
...   & ... & ... & ... & ... & ... & ... & ... \\ \hline
Row94 & 131 & 184 & 125 & ... & 0   & 238 & 161 \\ \hline
Row95 & 56  & 63  & 68  & ... & 118 & 0   & 93 \\ \hline
Row96 & 43  & 90  & 61  & ... & 113 & 97  & 0 \\
\bottomrule
\end{tabular}
\end{table}

\end{document}

%% file: Sec_1intro.tex
\section*{Background \& Summary} 

Cities serve as the primary hubs of social and economic activities~\cite{batty2013new,batty2013big,bettencourt2007growth,zhang2025metacity}, including manufacturing in factories, service provision in commercial establishments, trade in marketplaces, and technological innovation within research and development centers. To optimize efficiency, these activities are often spatially organized into functional zones~\cite{hillier1989social,montgomery1998making}, such as industrial zones for manufacturing and residential zones for housing. This spatial organization necessitates daily commuting between residential and workplace regions. Such critical behavior is typically captured in form of commuting Origin-Destination~(OD) flows, data that record population movements between residential and workplace regions~\cite{rong2023interdisciplinary,liu2020learning,iqbal2014development,simini2021deep}. Furthermore, commuting OD flows reflect peak mobility patterns during morning and evening rush hours, which constitute a significant component of urban dynamics~\cite{novaco2009commuting,dong2022universality}. Building on this, commuting OD flows are integral to advancing United Nations Sustainable Development Goals~(SDGs), and our systematic validation~(in Section of Technical Validation) for different goals demonstrates their reliability for sustainability:
\begin{itemize}[leftmargin=*]
    \item \textbf{Urban Planning~(SDG-11):} Commuting OD flows reveal inefficient long-distance travel patterns, enabling planners to optimize zone allocation and reduce travel costs~\cite{hamilton1982wasteful,banister2008sustainable,newman1998sustainability}. 
    \item \textbf{Urban Resilience~(SDGs-9,11):} OD flow patterns identify critical commuting corridors and vulnerable urban areas, working as a basis indicator for cities to develop targeted resilience strategies against disruptions~\cite{haraguchi2022human,ribeiro2019urban,mistry2021inferring}.
    \item \textbf{Transportation~(SDG-9):} OD flow data serves as the foundation for travel demand modeling and traffic simulations, providing insights into infrastructure loads and supporting strategies to mitigate traffic congestion~\cite{imai2021origin,cats2014dynamic,litman2012evaluating,newman1998sustainability}.
    \item \textbf{Public Health~(SDG-3):} Daily commuting OD patterns reveal high-risk transmission routes and crowded workplace regions, critical for epidemic control strategies~\cite{mistry2021inferring,balcan2009multiscale,kraemer2020effect}.
    \item \textbf{Energy Use~(SDG-7) and Environmental Protection~(SDG-13):} Commuting via different modes of transportation depends on a range of energy sources, spanning traditional fossil fuels to cleaner and more sustainable alternatives~\cite{xu2018planning,powell2022charging}. Analyzing OD flows provides critical insights into evaluating and managing energy demands, which is essential for minimizing the environmental impact of urban areas. Moreover, such analyses play a pivotal role in climate change mitigation by informing strategies to reduce carbon emissions~\cite{schwanen2021achieving,zeng2024estimating}.
\end{itemize}

However, global-scale commuting OD flow data covering diverse urban environments remains unavailable, hindering the universality and comprehensiveness of studies on urban dynamics and sustainable development. This limitation has resulted in existing works focusing on a small number of cities within specific urban contexts~\cite{karimi2020origin,pourebrahim2018enhancing,pourebrahim2019trip,liu2020learning,yao2020spatial,lenormand2015influence,lenormand2016systematic,rong2023complexity,rong2023goddag,simini2021deep}, leading to varied and context-dependent conclusions. The primary obstacle lies in the significant challenges associated with data collection. Traditionally, there are two main approaches to collecting commuting OD flows. The first is conducting travel surveys, which, while accurate, are time-consuming and expensive due to extensive population coverage required~\cite{rong2023goddag,rong2023interdisciplinary,axhausen2002observing,schuessler2009processing}. The second involves using location records collected from mobile phone usage~\cite{iqbal2014development,calabrese2011estimating,imai2021origin,alexander2015origin,toole2015path}, which is more cost-effective but faces barriers such as commercial conflicts among mobile service providers and privacy concerns, severely limiting global availability. Consequently, collection-based methods for acquiring commuting OD flows on a worldwide scale remain impractical and currently infeasible.

In this work, we present a large-scale commuting OD flow dataset for global cities, generated using deep graph generative models based on public data like population counts, demographics, satellite imagery, and POIs~(points of interest). The generation involves two processes: urban region profiling and commuting OD flow generation. First, we use public data to characterize urban regions' functions and attributes. Second, we model commuting flows as a weighted directed graph, where regions are nodes and flows are edges, and use graph generative models~\cite{rong2023complexity,rong2024large} to generate flows based on spatial characteristics. This data-driven approach enables realistic flow generation for specific urban spaces. We collect data for 1,625 cities across 179 countries and 6 continents, as shown in Figure~\ref{fig:global_cities}, enabling research on urban dynamics and supporting sustainable policy development globally.

We systematically validate our dataset from five perspectives to ensure its accuracy and robustness: (1) regression analysis based on mobile location data (correlation coefficients 0.74-0.82 for various cities); (2) error analysis against Census data (CPC=0.572, significantly higher than previous works); (3) geospatial visualization analysis in cities across Asia, Latin America, and Africa; (4) downstream tasks for predicting work population (R²>0.84 for all test cases) and carbon emission (correlation=0.66); and (5) attribute reproduction in scale-free commuting networks. The results demonstrate that our dataset is both reliable and valuable.

%% file: Sec_2method.tex
\section*{Methods} 
In this section, we provide a detailed description of the methodology used to construct the dataset. The construction pipeline, illustrated in Figure~\ref{fig:pipeline}, comprises three main steps. First, we select the cities to be included in the dataset and divide them into urban regions. Next, we profile these urban regions by extracting features from multi-source global public data using specific feature extraction techniques. Finally, we generate commuting OD flows between the urban regions using graph generative models.

\subsection*{Determining City Boundaries and Geographic Units}
To comprehensively cover cities and their geographic boundaries across diverse urban environments worldwide, we integrated multiple datasets to select cities with detailed boundary information. Specifically, we began by collecting cities from the WhosOnFirst~(WOF) database, a global open-source gazetteer of places. This data is available in GeoJSON format at \url{https://whosonfirst.org/}. From this dataset, we selected all cities with explicitly defined boundaries represented by polygons. To further enhance coverage, we supplemented our dataset with cities from the GUI dataset provided by Han et al.~\cite{han2024gui}, ensuring no duplication of cities already included in the WOF database. Through this integration process, we compiled a comprehensive dataset containing a total of 1,625 cities, each with detailed geographic boundary information.

After determining the cities and their boundaries, we divided each city into individual urban regions using a unified, adaptive grid-based method commonly employed in modeling urban dynamics~\cite{simini2021deep,wang2019origin,zhang2017deep,zheng2014urban,rong2023interdisciplinary}. To balance detail and computational efficiency, we dynamically adjusted the grid size for each city based on its geographic extent. This approach enables finer resolution for smaller cities and coarser grids for larger urban areas. The grid size was determined by the city's bounding box, defined by the distance between its minimum and maximum latitude and longitude. Specifically, the grid size was set to 5\% of the city's total boundary length, with an upper limit of 5 kilometers and a lower limit of 500 meters. This configuration ensures that the resulting urban regions are sufficiently detailed to capture the city's structure while remaining computationally manageable. Additionally, avoiding excessively small grids minimizes the impact of randomness and noise. Figure~\ref{fig:region_division_cases} illustrates examples of the region division for various cities, demonstrating the adaptability and effectiveness of this approach.

\subsection*{Multi-source Global Public Data for Profiling Urban Regions}
We utilize a variety of globally available public data sources, each rigorously validated for their effectiveness in describing the functional characteristics of urban regions, to generate commuting OD flows. Specifically, we use the following data sources:
\begin{itemize}[leftmargin=*]
    \item \textbf{Population and Demographics}: WorldPop~\cite{tatem2017worldpop} provides fine-grained global population counts and demographic data at a 100m resolution. Using this dataset, we calculate the total population and demographic structure (age and gender distributions) within each urban region. This information is essential for estimating the generation of population flows~\cite{barbosa2018human,zipf1946p,lenormand2016systematic}.
    \item \textbf{Satellite Imagery}: Satellite imagery offers a 2D bird's-eye view of the Earth's surface, capturing objects and structures indicative of human activities in urban areas. We use high-resolution imagery~(10m) provided by Esri World Imagery~\cite{esri_world_imagery}. For each urban region, we download image tiles within its boundaries and remove pixels located outside the defined region. Semantic features related to urban functional characteristics are then extracted using the vision encoder from RemoteCLIP~\cite{liu2024remoteclip}.
    \item \textbf{Points of Interest~(POIs)}: POIs refer to specific locations associated with human activities, and they are proven to be effective in characterizing urban functional traits~\cite{yuan2012discovering}. We obtain POIs from OpenStreetMap~(OSM)~\cite{OpenStreetMap} and filter them by type within each urban region's boundary. This process results in a vector representing the number of POIs for each type, which helps profile the functional characteristics of the region.
\end{itemize}

From each data source, we extract a feature vector for every urban region. These feature vectors are then concatenated to form the final semantic representation of each region, resulting in a 1,094-dimensional vector. This vector comprises 36 dimensions derived from population and demographic data, 1,024 dimensions from satellite imagery, and 34 dimensions from points of interest (POIs). The resulting feature vector serves as the node feature and acts as the conditional input for the graph generative model.

\subsection*{Generating Commuting OD Flows via Generative Graph Diffusion Models}
Using the extracted semantic features of urban regions as node-level inputs, we employ graph diffusion models to generate commuting OD flows, represented as weighted edges between nodes. Specifically, we adopt WEDAN~\cite{rong2024large}, a state-of-the-art model specifically designed for the generation of commuting OD flows. WEDAN has been demonstrated to outperform other methods for this task. As the primary focus of this work is to provide a synthetic dataset of commuting OD flows, we do not delve into the technical details of WEDAN. Seeking a deeper understanding of WEDAN's design and the principles of diffusion models can be referred to prior works~\cite{ho2020denoising,vignac2022digress}. For completeness, we provide a brief overview of WEDAN's utilization in this study.

WEDAN is a graph diffusion-based model that generates the weighted edges of a graph, representing commuting OD flows, conditioned on the node features. The framework of WEDAN comprises two primary processes: forward diffusion and reverse denoising. The forward diffusion process is used to prepare the training data by progressively adding small amounts of noise to the edge weights~(commuting OD flows) until the data is entirely disrupted and reduced to pure noise. The reverse denoising process then iteratively reconstructs the original data from the noise, guided by the node features. During each iteration of the reverse process, a portion of the noise is removed from the data produced in the previous iteration. The partially denoised data is then fed into the next iteration. This process continues until the noise is fully removed, resulting in data that conforms to the underlying data distribution. A key aspect of WEDAN is that the generation process is guided by the node features, which encapsulate the semantic representations of urban regions. This ensures that the generated commuting OD flows align closely with real-world urban dynamics.

%% file: Sec_3data.tex
\section*{Data Records} %

The dataset is available on the Figshare repository~\cite{rong2025figshare}, and a live version, which will be maintained and updated, is hosted on the GitHub repository https://github.com/tsinghua-fib-lab/WorldCommuting-OD. To support the research community with a comprehensive dataset, we provide data for each city, including the city boundary, region divisions, and commuting OD flows between urban regions. For ease of use, all data for each city is organized into a dedicated folder named after the city, ensuring straightforward navigation and access.

\subsection*{City Boundary and Region Division}
The data for city boundaries and region divisions is provided in Shapefile format. Each city has a corresponding Shapefile containing the geometry of its regions, organized as either Polygon or MultiPolygon geometry types. As shown in Table~\ref{tab:region_division_cases}, which is an example of the city of Athens, Greece, these Shapefiles can be loaded into Python as GeoDataFrames. Each Shapefile includes three key columns:
\begin{itemize}[leftmargin=*]
    \item \textbf{Index}: A unique identifier for each region. 
    \item \textbf{Locations}: The x-y coordinates of the grid region. As shown in Figure~\ref{fig:od_flow_vis_example}, based on the boundary box of the city, the whole city is divided into grids and remains only the regions within the boundary of the city. It worth noting that the x-y coordinates are different from the column and row indices in the commuting OD flow matrix, where the indices are the unique identifiers of the urban regions.
    \item \textbf{Geometry}: The boundary of urban regions, which are organized as either a Polygon or MultiPolygon. The Polygon consists of a list of Points, and the MultiPolygon consists of a list of Polygons. The start Point and end Point of each Polygon are the same, which ensures the continuity of the boundary. The geometry is constructed by the Shapely Python package.
\end{itemize}
A notable feature of the Shapefile format is that each region is assigned a fixed index, which serves as a consistent and unique identifier across different platforms and tools. For instance, when a Shapefile is loaded into Python using GeoPandas, it is read as a GeoDataFrame, with each row corresponding to a specific region and the index preserved as its unique identifier. Similarly, when imported into GIS software such as ArcGIS, the regions retain the same fixed index, ensuring consistency in region identification. This unique indexing system is critical for spatial analysis and data management, as it allows users to consistently track and reference individual regions within a city, regardless of the tool or platform being used.

\subsection*{Commuting OD Flows}
The commuting OD flows are provided in matrix form and stored as a NumPy array, as illustrated in Table~\ref{tab:od_flow_matrix}. In this matrix, the element at position $[i,j]$ represents the number of people residing in the $i$-th region of the city who commute to the $j$-th region for work. The indices $i$ and $j$ correspond to the unique identifiers~(indices) of the urban regions within the city's boundary and region divisions, as defined in the associated Shapefile. The matrices contains commuting OD flows between all urban regions within the specific cities' boundary. We visualize the commuting OD flows in three representative cities in Figure~\ref{fig:od_flow_3example} to give a intuitive physical representation of the commuting OD flows.

These indices are consistent with those in the Shapefile, where each region is assigned a fixed index that serves as its unique identifier. As a result, the region indices in the commuting OD flow matrix can be directly mapped to the corresponding urban regions defined in the Shapefile. This alignment facilitates seamless integration and spatial analysis, enabling the flow of people to be visualized or analyzed in relation to the city's geographic boundaries and region divisions.

For better understanding of the structure of our dataset, we provide an example using the city of Athens, Greece, a significant urban center in Europe. The city is divided into 96 grid-based urban regions, as detailed in Table~\ref{tab:region_division_cases}. Commuting OD flows between these regions are represented in a 96x96 matrix, which is provided in Table~\ref{tab:od_flow_matrix}. Additionally, Figure~\ref{fig:od_flow_vis_example} visualizes the commuting OD flows across the urban regions, with each region's index clearly marked on the map for reference. This example highlights the spatial granularity and structure of the dataset, facilitating analyses of intra-city commuting patterns and their implications for urban studies.

%% file: Sec_4eval.tex
\section*{Technical Validation}
We conduct a comprehensive technical validation of the generated commuting OD flows to ensure their accuracy and reliability. This validation encompasses evaluations across five aspects: (1) regression analysis based on mobile location data; (2) error analysis of generated flows against UK census data; (3) geospatial visualization analysis; (4) downstream tasks for predicting population distribution during work hours; and (5) attribute reproduction in scale-free commuting networks.

\subsection*{Regression Analysis Based on Mobile Location Data}
Existing studies have utilized mobile location data, such as CDRs and CNAs, to estimate commuting flows~\cite{gundlegaard2016travel,pan2006cellular,yang2021detecting}. For our validation, we collected mobile location data from various sources for representative cities worldwide, including Beijing and Shanghai in China, Rio de Janeiro in Brazil, and Senegal in Africa. We compared the generated commuting OD flows with commuting flows estimated from the mobile location data by calculating Spearman's rank correlation coefficient~(SCC) between the two datasets. As shown in Figure~\ref{fig:regression_analysis}, the SCC values for Beijing, Shanghai, Rio de Janeiro, and Senegal are 0.741, 0.641, 0.816, and 0.477, respectively. These results indicate a high degree of consistency between the generated commuting OD flows and the mobile location data. It is important to emphasize that while strong correlations validate the quality of the generated data, weaker correlations do not necessarily indicate poor quality. Such discrepancies may stem from limitations or biases inherent in the reference data. For instance, the relatively low SCC value of 0.477 for Senegal can likely be attributed to the lower coverage of mobile phone usage in Senegal compared to other cities.

\subsection*{Error Analysis of Generated Flows Against Census Data in the UK and US}
Thanks to the availability of comprehensive commuting OD flow data from national censuses in the UK and US, this provides a valuable benchmark for evaluating model performance. We leverage these data to conduct quantitative error analysis, offering a reliable measure for assessing the quality of the generation. Specifically, we utilize three metrics from two perspectives to quantitatively evaluate the deviation between the generated and census data: (1) common parts of commuting~(CPC), (2) root mean square error~(RMSE), and (3) normalized root mean square error~(NRMSE). The calculation of these metrics is detailed as follows.
\begin{equation}
    \begin{split}
    \begin{aligned}
        & \mathrm{RMSE} = \sqrt{\frac{1}{|\textbf{F}|} \sum\nolimits_{r_i,r_j \in \mathcal{R}} ||\textbf{F}_{ij} - \hat{\textbf{F}}_{ij}||_2^2}, \\
        & \mathrm{NRMSE} = \frac{\mathrm{RMSE}}{\sqrt{\frac{1}{N^2} \sum\nolimits_{r_i, r_j \in \mathcal{R}} || F_{ij} - \bar{F}_{ij} ||_2^2}}, \\
        & \mathrm{CPC} = \frac{2 \sum_{r_i, r_j \in \mathcal{R}} \min(\mathbf{F}_{ij}, \hat{\mathbf{F}}_{ij})}{\sum_{r_i, r_j \in \mathcal{R}} \mathbf{F}_{ij} + \sum_{r_i, r_j \in \mathcal{R}} \hat{\mathbf{F}}_{ij}}, \\
    \end{aligned}
    \end{split}
\end{equation}
where $\textbf{F}$ and $\hat{\textbf{F}}$ are the generated and census OD flows, respectively, $\mathcal{R}$ is the set of urban regions, $\textbf{P}_{\mathbf{F}}$ and $\textbf{P}_{\hat{\mathbf{F}}}$ are the probability distributions of $\textbf{F}$ and $\hat{\textbf{F}}$, respectively, and $\textbf{KL}$ is the Kullback-Leibler divergence.

We calculate the CPC, RMSE, and NRMSE between the generated and census data as 0.572, 47.91, and 0.927, respectively, which indicates that the generated commuting OD flows are highly consistent with the census data. This is a very high level in the perspective of commuting OD flow generation where no OD flows are provided as a reference. Previous studies in this field have reported a CPC of around 0.3~\cite{simini2021deep}, especially for the harder cases of cross-continent performance.

Furthermore, we present the distribution of CPC values across cities with varying numbers of urban regions in Figure~\ref{fig:cpc_distribution_size}. The results show that as city size increases, the CPC distribution transitions from being uniformly spread to converging around a value of 0.55. This indicates that as city size grows, the generated OD flow CPC values become increasingly stable. This trend suggests that our method produces more consistent results for larger cities, whereas smaller cities exhibit greater variability, likely due to increased noise. Additionally, we provide the spatial distribution of CPC values for two representative cities, Chicago and London, in Figure~\ref{fig:cpc_distribution_spatial}. As shown, CPC values are higher in areas with greater OD flow. This implies that the generated commuting OD flows achieves greater precision in regions with higher flow volumes, enhancing the utility of the generated data. These high-flow areas often correspond to critical urban regions, making the data more suitable for applications targeting key urban zones.

\subsection*{Geospatial Visualization Analysis}
To intuitively validate the generated commuting OD flows, we visualize these flows on city maps and compare them with other related data sources, such as mobile location data. Cities from various continents, including China, Latin America, and Africa, are used to assess the spatial consistency between the generated commuting OD flows and the reference data sources. The comparison results are presented in Figure~\ref{fig:geovisualization_beijing},\ref{fig:geovisualization_shanghai},\ref{fig:geovisualization_rio}, and~\ref{fig:geovisualization_senegal}. As shown, the generated commuting OD flows exhibit strong spatial consistency with the reference data. Key features such as commuting hotspots and city centers are accurately captured. These results demonstrate that the generated commuting OD flows are reliable and suitable for further analysis and applications.

We also provide the geospatial visualization of commuting OD flows of representative polycentric cities in Figure~\ref{fig:geovis_polycentric_cities}. As we can see that the generated commuting OD flows are able to capture the polycentric structure of the cities, with high commuting flows between the central city and the satellite cities. This demonstrates that the generated dataset is comprehensive and adaptable to different urban structures.

\subsection*{Downstream Tasks in Multiple Urban Applications and Corresponding SDGs}
To further validate the generated commuting OD flows, we perform downstream tasks to predict population distribution during work hours. This is achieved by integrating the nighttime residential population distribution with the commuting OD flows to model daytime transitions. Specifically, we use nighttime population distribution data from WorldPop as the baseline and incorporate transitions from residential locations to workplace locations during commuting hours as the variation to derive the population distribution during work hours. To further enhance the validation, we compare the predicted daytime population distribution against real-time population heatmap data from Baidu Map's Huiyan urban population geography platform. The heatmap data is scaled to approximate the full population distribution during work hours using the method proposed by Xu et al.~\cite{xu2016context}. As illustrated in Figure~\ref{fig:population_prediction}, the results demonstrate that our generated commuting OD flows accurately predict population transitions from nighttime residential periods to daytime work periods, with coefficients of determination (R$^2$) of 0.95, 0.89, 0.84, and 0.87 for Chengdu, Guangzhou, Jinan, and Qingdao, respectively. These findings provide strong evidence for the quality and reliability of the generated commuting OD flows.

To further validate the practical utility of our generated commuting OD flows, we apply them to estimate urban transportation carbon emissions as a downstream task~(\textbf{SDGs-9,13})~\cite{zeng2024estimating}. Specifically, we first calculate commuting distances based on OD matrices and regional centroids, incorporating city-specific detour factors to account for different road network characteristics. We then combine these with modal shares (60\% private cars, 35\% public transit, and 5\% cycling/walking) and their respective emission factors (171, 104, and 0 gCO$_2$/km) to compute the overall transportation carbon emissions for each city. As shown in Figure~\ref{fig:carbon_estimation}, we compare our estimates with actual measurements from Carbon Monitor Cities~\cite{huo2022carbon}. From the results, our estimates demonstrate strong correlation with the measured data (correlation coefficient of 0.66). This carbon emission estimation method based on commuting OD flows not only validates the reliability of our generated data but also provides a novel approach for rapid assessment of urban transportation carbon emissions. This approach is particularly valuable for cities lacking detailed carbon emission monitoring data, offering a practical way to estimate their transportation-related carbon footprint.

We also employ them to analyze urban resilience (\textbf{SDGs-9,11}). Specifically, we develop a comprehensive resilience assessment framework that incorporates multiple dimensions of urban networks. We first calculate the network connectivity based on OD matrices, measuring both the strength and distribution of commuting flows. We then evaluate system vulnerability by simulating node failures and quantifying their impacts on the overall network performance. Additionally, we assess network redundancy through alternative path analysis and recovery potential based on the distribution of critical nodes~\cite{meerow2016defining,reggiani2013network,derrible2017urban,sun2018vulnerability,batty2012origins}. As shown in Figure~\ref{fig:resilience_groups_analysis}, we analyze the relationship between city size and urban resilience for 21 major global cities. The results reveal two distinct patterns of urban resilience: high-resilience cities~(R²=0.679) and low-resilience cities~(R²=0.272). Notably, several megacities like Shanghai, Chicago, and Paris exhibit high resilience despite their large populations, suggesting that well-planned urban structures can maintain system resilience even under the pressure of massive population. Furthermore, there is a significant correlation between city size and urban resilience, with larger cities generally exhibiting lower resilience, which can clearly be profiled by our generated commuting OD flows.

\subsection*{Attribute Reproduction in Scale-free Commuting Networks}
Existing studies examining the commuting behavior as networks from the perspective of complex networks have established that these networks are scale-free, meaning the number of commuters between two regions follows a power-law distribution. To validate this property in the generated commuting networks, we analyze and visualize the distribution of commuting flows. As shown in Figure~\ref{fig:scaling_analysis}, the distribution of the generated commuting flows closely follows a power-law distribution, indicating that the generated commuting networks exhibit the scale-free property. This result further validates the quality of the generated commuting OD flows, demonstrating their consistency with real-world commuting networks.